\documentclass[fleqn,usenatbib]{mnras}

\usepackage{newtxtext,newtxmath}

\usepackage[T1]{fontenc}
\usepackage{ae,aecompl}

\usepackage{graphicx}	
\usepackage{amsmath}	
\usepackage{subfigure}
\usepackage{xcolor}

\def\mnras{MNRAS}
\def\apj{ApJ}
\def\apjl{ApJL}
\def\apjs{ApJS}
\def\aap{A\&A}

\title{Multi-wavelength detectability of isolated black holes in the Milky Way} 

\author[F. Scarcella et al.]
{
Francesca Scarcella,$^{1}$\thanks{E-mail: francesca.scarcella@uam.es}
Daniele Gaggero,$^{1}$\thanks{E-mail: daniele.gaggero@uam.es}
Riley Connors,$^{2}$
Julien Manshanden,$^{3}$
\newauthor
Massimo Ricotti,$^{4}$
Gianfranco Bertone$^{5}$
\\
$^{1}$ Instituto de F\'isica Te\'orica UAM-CSIC, Campus de Cantoblanco, E-28049 Madrid, Spain\\
$^{2}$ Cahill Center for Astronomy and Astrophysics, California Institute of Technology, Pasadena, CA 91125, USA\\
$^{3}$ II. Institut für Theoretische Physik, Universität Hamburg, 22761 Hamburg, Germany\\
$^{4}$ Department of Astronomy, University of Maryland,
College Park, MD 20740, USA\\
$^{5}$ Gravitation Astroparticle Physics Amsterdam (GRAPPA), Institute for Theoretical Physics Amsterdam,
and Delta Institute for Theoretical Physics,\\ University of Amsterdam, Science Park 904, 1098 XH Amsterdam, The Netherlands\\
}

\date{Accepted XXX. Received YYY; in original form ZZZ}

\pubyear{2020}

\begin{document}
\label{firstpage}
\pagerange{\pageref{firstpage}--\pageref{lastpage}}

\maketitle

\begin{abstract}
Isolated black holes in our Galaxy have eluded detection so far. We present here a comprehensive study on the detectability of isolated stellar-mass astrophysical black holes that accrete interstellar gas from molecular clouds in both the local region and the Central Molecular Zone. We adopt a state-of-the-art model for the accretion physics backed up by numerical simulations, and study the number of observable sources in both the radio and X-ray band, as a function of a variety of parameters. We discuss in particular the impact of the astrophysical uncertainties on our prediction for the number of bright X-ray sources in the central region of the Galaxy. We finally consider future developments in the radio domain, and assess the potential of SKA to detect a population of astrophysical black holes accreting gas in our Galaxy.
\end{abstract}

\begin{keywords}
astroparticle physics -- black hole accretion -- ISM: jets and outflows
\end{keywords}

\section{Introduction}

A large population of Stellar-mass black holes (BHs) is believed to exist in our Galaxy: several works \citep{Shapiro:1983du,1998ApJ...496..155S,2017MNRAS.468.4000C} agree on an order-of-magnitude estimate of $\sim 10^8$ BHs produced by the collapse of massive stars in the Galaxy.
This large number of BHs has eluded detection so far, with the exception of $\simeq 60$ objects\footnote{\url{http://www.astro.puc.cl/BlackCAT/}} that are part of binary systems and accrete a significant amount of mass from a companion star: such systems are known as black-hole X-ray binaries (BH-XRBs) and appear as bright X-ray sources in the sky.

Here, we focus on {\it isolated} BHs, whose detection represents a major challenge in modern astronomy.
The problem has been discussed in several recent studies focused on the future detectability of the multi-wavelength radiation possibly emitted by these objects when they accrete matter from interstellar clouds either in the vicinity of the Sun \citep{Maccarone:2005,Fender:2013} or in the regions near the Galactic centre, as recently discussed in \citealt{Tsuna:2019kny,Tsuna:2018oqt}. 
The latter region seems particularly promising. In fact, an extensively studied complex of giant molecular clouds usually called {\it Central Molecular Zone} (CMZ) characterize a vast region that extends up to $\simeq 200$ pc away from the Galactic Centre, and provides a huge reservoir of molecular hydrogen. The high-density clumps within these region appear as ideal targets for astronomical searches of compact objects based on gas accretion.

In order to perform these analyses, it is crucial to develop an accurate modeling of {\it (i)} the accretion process of baryonic matter onto a compact object, in order to compute the expected accretion rate as function of the environmental conditions and the BH mass and speed; and {\it (ii)} the non-thermal spectrum of the radiation emitted by the accreted gas in different wavelengths, from radio all the way up to the hard X-ray band. 

Regarding the former aspect, a simplified model for the accretion physics is adopted in the aforementioned papers, based on a rescaling of the Bondi-Hoyle-Lyttleton (BHL) formula. 
However, the accretion problem exhibits a richer phenomenology. Radiation feedback plays a crucial role, and is expected to shape the surrounding environment in a remarkable way, forming a cometary-shaped ionized region around the compact object. A state-of-the-art treatment of accretion in the presence of radiative feedback, backed up by numerical simulations, was presented in a series of works (\citealt{Park:2010yh,Park:2011rf,SugimuraR:2020} and \citealt{Park:2012cr}, PR13 hereafter)  and showed a more complicated behaviour of the accretion rate with respect to the BH speed compared to the BHL scaling. This model features an increase at low velocities, and an asymptotic decrease to the BHL rate at large velocities.
The PR13 results were recently applied to compact object searches in the context of the quest for BHs of primordial origin \citep{Manshanden:2018tze}. 
However, a comprehensive study of this model related to the population of astrophysical BHs is still lacking.

As for the prescription for the multi-wavelength (radio/X-ray) non-thermal emission, the absence of any observational detection of isolated accreting stellar-mass BHs means any physical model one can develop carries large uncertainties. However, one can utilize the vast knowledge gathered in the studies of both accreting BH-XRBs  and accreting supermassive BHs to inform such a prescription. In particular, the {\it Fundamental Plane} relation between radio and X-ray emission \citep{2012MNRAS.419..267P} derived from the study of the aforementioned systems can be a precious tool to characterize the relation between radio and X-ray emission.

The aim of this paper is to characterize the detection prospect of a population of astrophysical black holes in view of the future development of radio and X-ray astronomy (with particular reference to the SKA experiment for the former), taking into account the state-of-the-art accretion formalism from PR13.

The paper is organized as follows: In Secs. \ref{sec:accretion} and \ref{sec:emission} we review the main aspects of accretion physics with particular focus on the PR13 model, and discuss the emission mechanisms that are most relevant in the radio and X-ray domain. Then, we present our methodology in Sec. \ref{sec:estimate} and \ref{sec:bhpopulation} and turn our attention to the astrophysical black holes in the vicinity of the Sun in Sec. \ref{sec:searchlocal} estimating the number of detectable X-ray sources. We then focus on the Central Molecular Zone in Sec. \ref{sec:searchcmz}, and present a comprehensive parametric study of the expected X-ray emission from astrophysical BHs in this region. Finally, in Sec. \ref{sec:SKA}, we present a forecast about the number of radio sources potentially detectable by SKA in association with the population of isolated astrophysical BHs accreting gas in the CMZ.

\section{Setting the stage: State-of-the-art accretion physics}
\label{sec:accretion}

In this section we describe in detail the accretion model introduced in PR13 and compare it to the textbook  Bondi-Hoyle-Lyttleton model.

\subsection{The Bondi-Hoyle-Lyttleton model}
\label{sec:BHL}

According to the BHL accretion model \citep{Hoyle1939,Bondi1944}, the rate of accretion onto an isolated compact object of mass $M$ moving at a constant speed ${\rm v}_\mathrm{BH}$ is given by:
\begin{equation}
\label{eq:bondi}
    \dot{M}_{\rm BHL} = 4 \pi \frac {(GM)^2 \rho} {({\rm v}_\mathrm{BH}^2 + c_{\rm s}^2)^{3/2}} \,\,\, ,
\end{equation}
where $\rho$ and  $c_{\rm s}$ are, respectively, the density and the sound speed that characterize the ambient medium and $G$ is the gravitational constant. 
This formula is usually re-scaled by introducing a suppression factor $\lambda$. \cite{Fender:2013} for instance, found that values larger than $\lambda=0.01$ are excluded, under realistic assumptions, by observations of the local region, where a significant population of isolated BHs should be present. \cite{Tsuna:2019kny} and \cite{Tsuna:2018oqt} consider values between $\lambda=0.1$ and $\lambda=10^{-3}$. The $\lambda$ parameter may effectively capture the outflow of material that is expelled from the Bondi sphere, as suggested by several authors \citep{Blandford:1999}. Its introduction is also supported by the non observation of a large population isolated neutron stars in the local region \citep{Perna:2003} and the studies of nearby AGNs \citep{Pellegrini:2005} as well as the supermassive BH at the Milky Way centre, Sagittarius A* \citep{2003ApJ...591..891B}. In conclusion, it seems that the BHL accretion formula may overestimate the accretion rate by orders of magnitude, although the physical mechanism behind this deviation is still disputed. One such mechanism is introduced in PR13, as we discuss below.

\subsection{The Park-Ricotti model}

The hydrodynamical simulations performed in PR13 show that, when radiative feedback is taken into account, a cometary-shaped ionized region is formed around the BH as it moves through the interstellar medium.

The model proposed in the same work combines the BHL formula with the modelling of the ionization front to obtain an analytical formula in agreement with the results of the simulations. We review it here. 
First, BHL accretion is assumed to hold within the ionized region, so that the PR13 accretion rate can be written as:

\begin{equation}
\label{eq:ricotti}
    \dot{M}_{\rm PR13} = 4 \pi \frac {(GM)^2 {\rho}_{\rm in}} {({\rm v}_{\rm in}^2 + c_{\rm s, in}^2)^{3/2}} \,\,\, ,
\end{equation}
where $c_{\rm s, in}$ and ${\rho}_{\rm in}$ are the sound speed and density of the ionized medium, and  ${\rm v}_{\rm in}$ is its velocity relative to the BH. The sound speed $c_{\rm s, in}$ is a free parameter of the model (of the order of few tens of km/s).
Now Euler's equations can be applied to the ionization front to express ${\rho}_{\rm in}$ and ${\rm v}_{\rm in}$ in terms of the corresponding quantities referred to the external neutral medium: its sound speed $c_s$, its density $\rho$ and the relative velocity of the BH, ${\rm v}_\mathrm{BH}$.

We briefly discuss here the derivation of these relations. Assuming a one-dimensional steady flux, the jump conditions obtained by applying Euler's equations for mass and momentum conservation at the ionization front are: 
\begin{align}
    \label{eq:masscons}
     &{\rho}_{\rm in}{\rm v}_{\rm in}  = {\rho}\,{{\rm v}_\mathrm{BH}}  \,\,\, ,      \\[5pt]
     \label{eq:momcons}
    &{\rho}_{\rm in}({\rm v}_{\rm in}^2+{c}_{\rm s,in}^2)  =  {\rho}({\rm v}_{\rm BH}^2+{c}_{\rm s}^2) \,\,\, ,
\end{align}
where to obtain the second equation we have further assumed an ideal gas. These equations have the following solutions:
\begin{equation}
\begin{aligned}
    &\rho_{\rm in } = \rho_{\rm in }^{\pm} \equiv \rho_{\rm} \frac{{\rm v}_{\rm BH}^2+c_{\rm s}^2 \pm \sqrt{\Delta}}{2 \, c_{\rm s,in}^2 } \; , \qquad \Delta \equiv ({\rm v}_{\rm BH}^2+c_{\rm s}^2)^2 - 4 \,{\rm v}_{\rm BH}^2 \, c_{\rm s,in}^2 \label{eq:deltarho}\\
    &{\rm v}_{\rm in} = \frac{{\rho}_{\rm}}{{\rho}_{\rm in}}{\rm v}_{\rm BH}  \; ,
\end{aligned}
\end{equation}
for ${\rm v}_{\rm BH} \leq {\rm v}_{\rm D}$  or ${\rm v}_{\rm BH} \geq {\rm v}_{\rm R}$, with ${\rm v}_{\rm D}$ and ${\rm v}_{\rm R}$ the roots of $\Delta$. Since typically $c_{\rm s,in} \sim \mathcal{O}$ (10) km/s while $c_{\rm s,out} \sim \mathcal{O}$ (1) km/s, we have:
\begin{equation}
\label{eq: velocities}
\begin{aligned}
 &{\rm v}_{\rm R}
 \approx 2 c_{\rm s,in} \, , \\
 &{\rm v}_{\rm D}
 \approx \frac{c_{\rm s}^2}{2 c_{\rm s,in}} \; \ll 1 \rm{km/s} .
\end{aligned}
\end{equation}

In the high-velocity range, $ {\rm v}_\mathrm{BH} \geq {\rm v}_{\rm R}$, the correct solution is given by $\rho_{\rm in}^-$, while for $ {\rm v}_\mathrm{BH} \leq {\rm v}_{\rm D}$, the front is best described by by $\rho_{\rm in}^+$ \citep{Park:2012cr}. In the intermediate-velocity range, ${\rm v}_D < {\rm v}_{\rm BH} < {\rm v}_R$, no real common solution can be obtained for Eqs. \ref{eq:masscons} and \ref{eq:momcons}. Correspondingly, as velocity lowers below ${\rm v}_R$, the pressure wave building behind the ionization front detaches from it as a shock into the neutral material. Part of the flux of matter is deviated and the velocity of the gas beyond the shock is reduced to below ${\rm v}_D$. This implies on the one hand that Eq. \ref{eq:masscons}, stating the mass conservation through the front along the direction of displacement of the BH, is no longer valid. On the other hand, since the velocity past the shock is now lower than ${\rm v}_D$, the jump conditions at the ionization front can be solved. One should now consider, however, two sets of jump conditions associated to the two fronts: the shock and the ionization front. Park and Ricotti instead observed through simulations that, in this regime, the velocity inside the ionized region is ${\rm v}_{\rm in} \approx c_{\rm s, in} $. This relation, promoted to an equality, can be used together with Eq. \ref{eq:momcons} to compute the density $\rho_{\rm in}$. This way, for  ${\rm v}_D \leq {\rm v}_{\rm BH} \leq {\rm v}_R$, we obtain:
\begin{equation}
\begin{aligned}
    &\rho_{\rm in } = \rho_{\rm in }^0 \equiv \rho \frac{{\rm v}_{\rm BH}^2+c_{\rm s}^2 }{2 \, c_{\rm s,in}^2 } \;, \\
    &{\rm v}_{\rm in} = {c}_{\rm s,in} \, .
\end{aligned}
\end{equation}

Thus we have in summary: 
\begin{equation}
  \rho_{\rm in} \quad 
    = \quad
  \begin{cases}
    \quad
    \rho_{\rm in}^- \; , \qquad 
    & \, {\rm v}_{\rm BH} \geq {\rm v}_{\rm R} \; , \\[5pt]
    \quad  
    \rho_{\rm in}^0 \; , \qquad 
    & \, {\rm v}_{\rm D } < {\rm v}_{\rm BH} < {\rm v}_{\rm R} \; , \\[5pt]
    \quad
    \rho_{\rm in}^+ \; , \qquad
    & \, {\rm v}_{\rm BH} \leq {\rm v}_{\rm D } \; ,
  \end{cases}\\[20pt]
\end{equation}
and
\begin{equation}
   {\rm v}_{\rm in} \quad 
     = \quad
  \begin{cases}
    \quad
    \dfrac{\rho}{{\rho}_{\rm in}}{\rm v}_{\rm BH} \; , \qquad
    & \, {\rm v}_{\rm BH} \geq {\rm v}_{\rm R} \; , \\[5pt]
    \quad  
    c_{\rm s, in} \; , \qquad
    & \, {\rm v}_{\rm D } < {\rm v}_{\rm BH} < {\rm v}_{\rm R} \; , \\[5pt]
    \quad
    \dfrac{\rho}{{\rho}_{\rm in}}{\rm v}_{\rm BH} \;, \qquad
    & \, {\rm v}_{\rm BH} \leq {\rm v}_{\rm D } \; .
  \end{cases}\\[20pt]
\end{equation}

Plugging these equations back into Eq. \ref{eq:ricotti} finally gives the desired accretion rate expressed in terms of the BH speed and the properties of the neutral medium it is moving through.

\begin{figure}
\begin{subfigure}
  \centering
  \includegraphics[width=\linewidth]{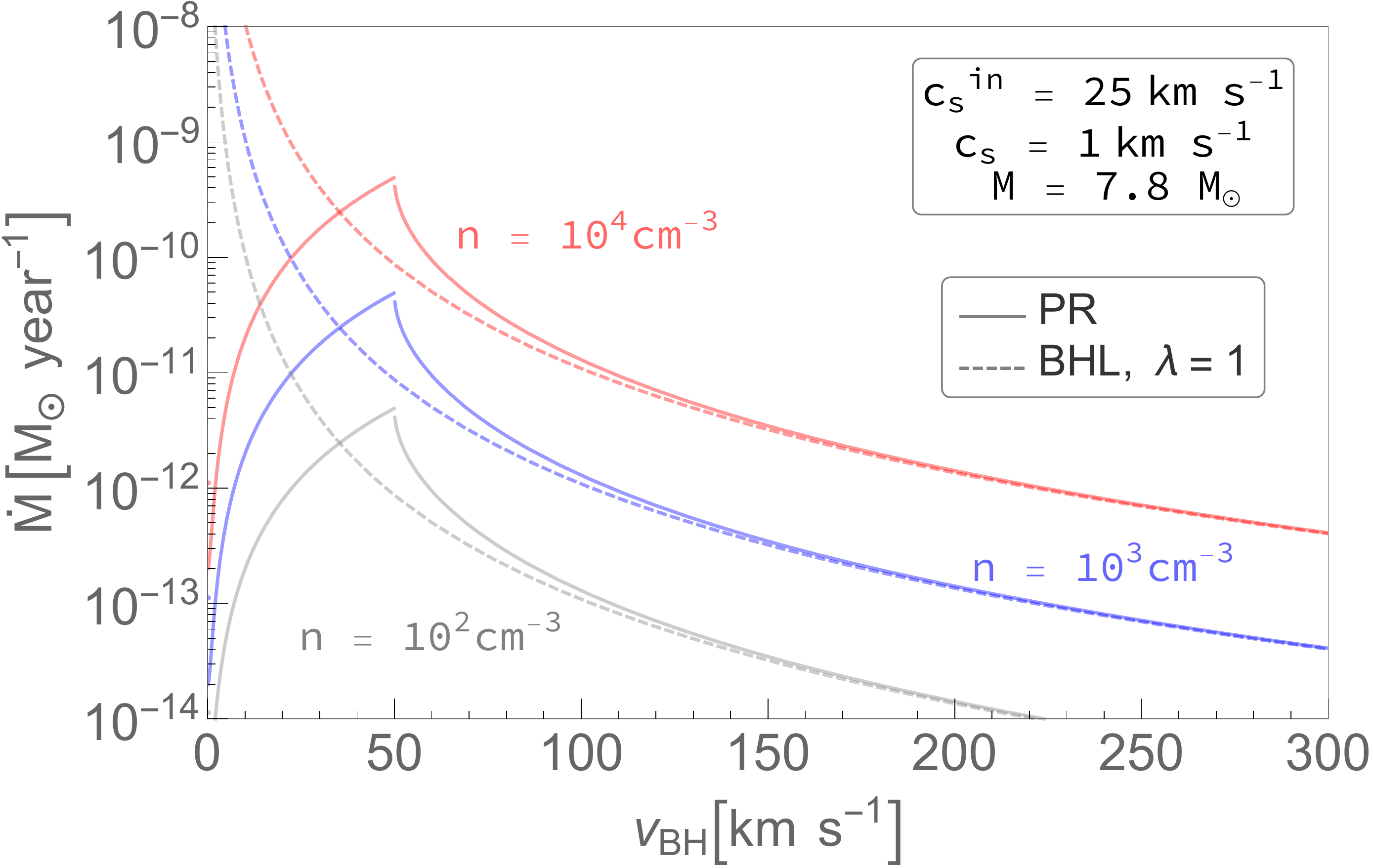}
\end{subfigure}
\begin{subfigure}
  \centering
  \includegraphics[width=\linewidth]{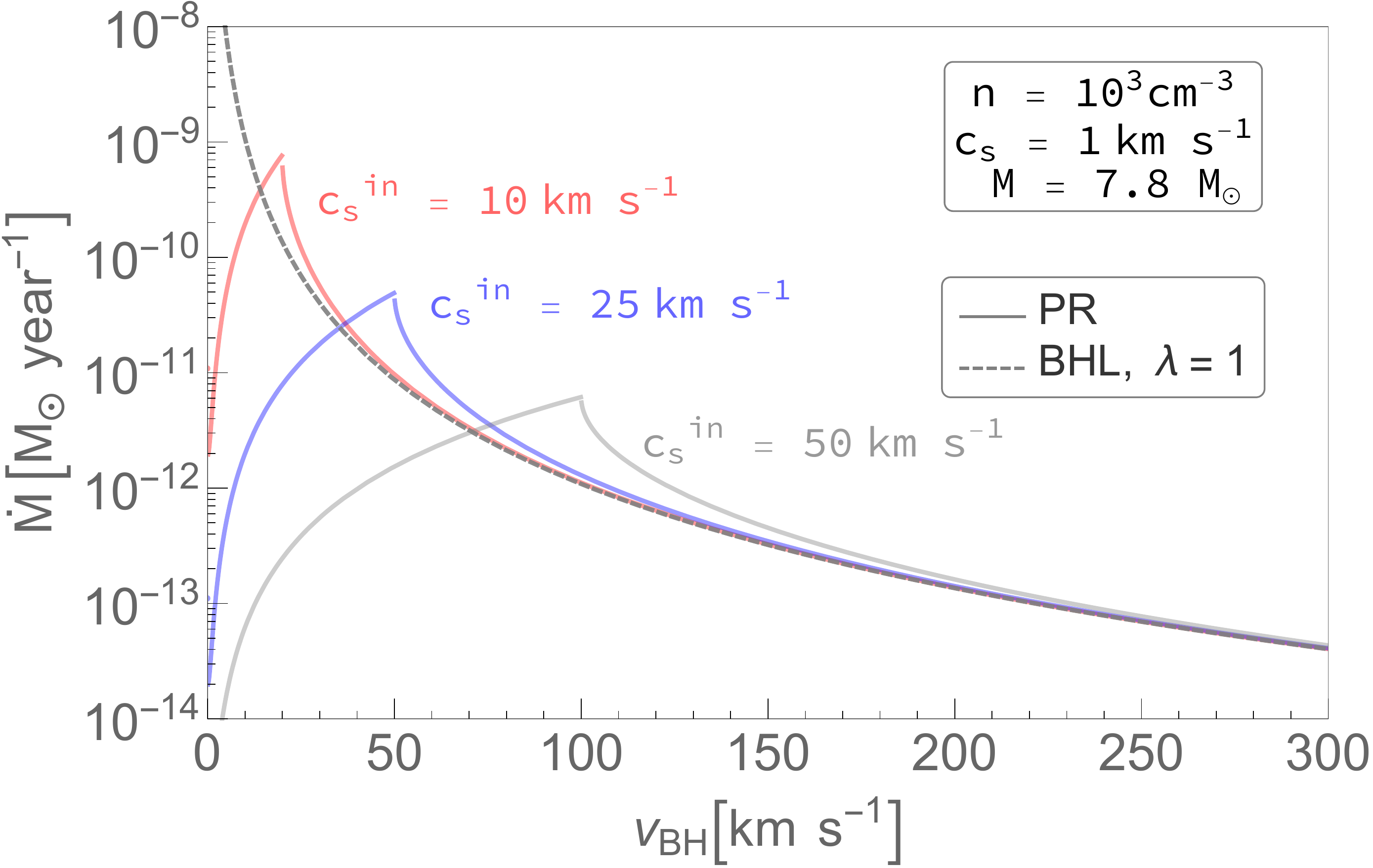}
\end{subfigure}
\begin{subfigure}
  \centering
  \includegraphics[width=\linewidth]{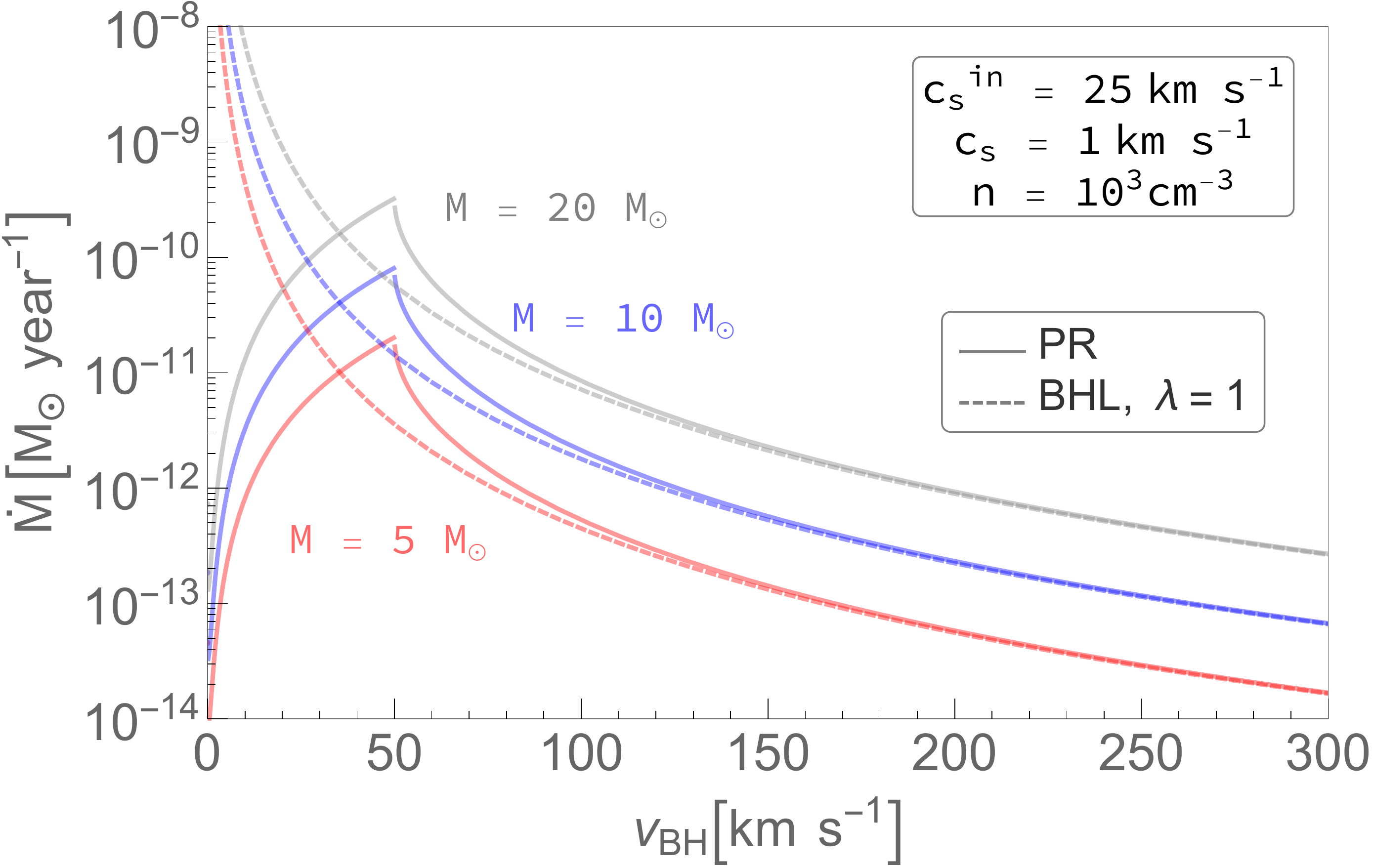}
\end{subfigure}
\caption{ {\it {\bf Accretion rate as a function of the BH speed.} We show the accretion rate obtained according to the PR13 and BHL models, as a function of the BH speed and other relevant parameters. For the BHL rate, we set the suppression factor $\lambda =1$, to allow for a more direct comparison. The to models agree at high velocity, but predictions differ by many orders of magnitude in the low velocity range. }}
\label{fig:accretion}
\end{figure}

Notice in particular that in the velocity range ${\rm v}_\mathrm{D} \leq {\rm v}_\mathrm{BH} \leq {\rm v}_\mathrm{R}$ we get:
\begin{align}
    \dot{M}_\mathrm{PR13} = \pi \frac{(GM)^2 \rho ({\rm v}_\mathrm{BH}^2 + c_\mathrm{s}^2)}{\sqrt{2} \, c_\mathrm{s,in}^5} \, ,
    \label{eq:intermediatePR13}
\end{align}
which increases quadratically with the BH velocity, in sharp contrast to the behaviour of the BHL rate, which decreases with velocity. This behaviour is the main feature introduced by the PR13 model, and is due to the formation around these objects of the aforementioned bow shock that deflects part of the gas away from the BH. The velocity dependence of the BHL rate is recovered in the high velocity regime, ${\rm v}_\mathrm{BH} > {\rm v}_\mathrm{R}$, where both rates present a $\propto {\rm v}_\mathrm{BH}^{-3}$ dependence. The complete velocity dependence of the PR13 rate is shown in figure \ref{fig:accretion}, for varied gas densities, BH masses, and sound speeds of the ionized region. For comparison, the BHL rate with $\lambda = 1$ is also shown. We can observe in this figure how the  BHL rate decreases monotonically with velocity, whereas the PR13 rate peaks at $ {\rm v}_\mathrm{BH} = {\rm v}_\mathrm{R} = 2 \, c_{\rm s, in}$ and is suppressed at lower velocity by the presence of the bow shock.

For $v < {\rm v}_{\rm D} $, the rate increases again. However, notice that this transition typically happens at velocities of $\approx 0.1$ km/s (see Eq. \ref{eq: velocities}), which are of little relevance for this work and not shown in figure.

The different velocity dependence of the PR13 rate compared to the BHL rate has important consequences when studying the emission properties of a BH population characterized by a given velocity distribution. According to the BHL prescription, the low-velocity tail of the population is the easiest to detect. Following PR13, the highest emissions are instead associated to BHs with intermediate velocities.

Furthermore, BHL can predict very high accretion rates if the speeds are low enough, which can easily lead to overshooting experimental bounds, while the highest rates predicted by PR13 are orders of magnitude smaller.

As a final remark, using Eq. \ref{eq:intermediatePR13} to express the peak of the PR13 rate in terms of the Eddington accretion rate $\dot{M}_\mathrm{Edd}$ :
\begin{align}
    \frac{\left. \dot{M}_\mathrm{PR13} \right\rvert_{{\rm v}_\mathrm{BH}={\rm v}_\mathrm{R}}}{\dot{M}_\mathrm{Edd}} \approx 10^{-4}
    \left( \frac{M_\mathrm{BH}}{10 \, M_\odot} \right) \left( \frac{\rho/m_p}{10^3  \, \mathrm{cm}^{-3}} \right) \left( \frac{c_{s,\mathrm{in}}}{25 \, \mathrm{km/s}} \right)^{-3}  \, ,
    \label{eq:PR13_rate}
\end{align}
one can see that the accretion rate will always be highly sub-Eddington in the range of BH masses $M_\mathrm{BH}$, gas densities $\rho$ and ionized sound speeds $c_{s,\mathrm{in}}$ we consider in this work. This is taken into account when defining the prescription for the luminosity in the next section.

\section{Emission mechanisms}
\label{sec:emission}

In order to translate the predicted accretion rates of a population of isolated BHs into a prediction of detectable sources we need estimates of the associated bolometric luminosity of a given source, and the Spectral Energy Distribution (SED). Our focus is on the X-ray and radio luminosity, $L_{\rm X}$ and $L_{\rm R}$, of the accreting BH. We can estimate both by considering the SED of the BH at varying accretion rates. We are particularly interested in emission from highly sub-Eddington accreting BHs (as predicted by Eq.~\ref{eq:PR13_rate}). Therefore in this section we describe our simple framework of the properties of sub-Eddington flows onto isolated BHs based on our prior understanding of known weak accretors in nature: Galactic BH-XRBs, and low-luminosity AGN. 

Radiatively Inefficient Accretion Flow (RIAF) models have spearheaded many studies regarding the emission mechanisms associated with such sub-Eddington accretion flows, with a focus both on Galactic BH-XRBs at low accretion rates, and the supermassive BH in the Galactic center, Sgr~A* \citep{Narayan:1994,yqn03}. 

The regularly cited RIAF models which first considered the emission processes in inefficiently radiating accreting BHs are classed as advection-dominated accretion flows (ADAFs; \citealt{Narayan:1994,Esin1997}). In the ADAF model, Bremsstrahlung radiation dominates the observed spectrum at the lowest accretion rates, and thus the spectrum has some curvature and its peak energy depends on the thermal gas temperature. As the accretion rate and gas density increase, Inverse Compton (IC) scattering begins to dominate (likely via SSC, i.e., Synchrotron photons become scattered), though if densities remain low enough, the spectrum will still show come curvature. At higher accretion rates, the IC scattering process is more efficient, resulting in a power-law-like spectrum in the X-ray band. 

The key thermodynamic property of RIAFs is the inefficient cooling of ions due to the Coulomb decoupling of electrons and ions in the accreting, low-density plasma. Due to this decoupling, such flows are likely well described by a two-temperature electron-ion plasma, with only electrons radiating via Bremsstarhlung, Synchrotron, and inverse Compton scattering \cite{Esin1997}. The models, as well as a plethora of observational evidence, show that such flows display radiative inefficiency, with $\eta=0.1\dot{M}/\dot{M}_{\rm crit}$, where $\eta$ is defined by:
\begin{equation}
    L = \eta \dot{M} ,
\end{equation}
and $\dot{M}_{\rm crit}$ is the accretion rate below which we have a RIAF. We assume $\dot{M}_{\rm crit}$ is the accretion rate corresponding to $1\%$ of the Eddington luminosity with $\eta_{\rm Edd }= 0.1$ \citep{Fender:2013}.The bolometric luminosity $L$ in such inefficient states is thus given by

\begin{equation}
    L = 0.1 \frac{\dot{M}^2}{\dot{M}_{\rm crit}},
\end{equation}

thus exhibiting a quadratic scaling $L\propto \dot{M}^2$ \citep{Esin1997}, as opposed to the linear behaviour observed at higher efficiency. 

A significant fraction of this bolometric luminosity is typically assumed to fall in the X-ray band. The $L_X/L$ fraction is usually assumed to be $\simeq 30\%$ \citep{Fender:2013}. 

Radiatively inefficient Galactic BH-XRBs also almost ubiqitously launch outflows. Such outflows are typically in the form of steady, self-absorbed relativistic jets \citep{Fender:2001}. Ballistic, transient jets are also observed during spectral state transitions at higher accretion rates. Such jets are typically identified via their radio emission, and are shown to be present during quiescence, at luminosities below $10^{-8}~L_{\rm Edd}$ \citep{Plotkin2015}. These steady jets exhibit a flat-to-inverted spectrum from radio through to IR frequencies---a consequence of self-absorbed synchrotron emission through the optically-thick regions of the jet (see, e.g., \citealt{Markoff2005}). 

In addition, multiple studies have now established that the luminosity of such radio-emitting jets scales in a consistent and predictable way with the X-ray emitting plasma in the inner regions of the flow. This scaling, $L_{\rm R} \propto L_{\rm x}^{0.7}$ \citep{Corbel2000,Corbel2003,Gallo2003,Corbel2008,MillerJones2011,Gallo2014}, is known as the radio-X-ray correlation, and applies to Galactic BH-XRBs (though other compact objects have been tracked in this phase space, showing similar but distinct trends of their own). Therefore, by using this analogy between isolated BHs and their low-accretion rate counterparts in binaries, we can infer that isolated BHs will similarly launch these steady jets, and thus their radio and X-ray luminosities will scale according to this radiatively inefficient track we observe in Galactic BH-XRBs. 

However, in order to invoke a mass scaling and capture low-luminosity accretion onto a mass-variable population of isolated BHs, one must refer to the established connection between Galactic BH-XRBs and AGN---the Fundamental Plane of Black Hole Activity (FP; \citealt{Merloni:2003,F04,Plotkin:2012}). The FP is, in a sense, an extension of the radio-X-ray correlation discovered in Galactic BH-XRBs to their supermassive counterparts. The FP is an empirical, parameterized relation between the X-ray luminosities, radio luminosities, and masses of hard state Galactic BH-XRBs with steady jets, and AGN of types which display similar X-ray emission characteristics. These include low-luminosity AGN (LLAGN), low-ionization nuclear emisssion-line regions (LINERS), Faranoff-Riley type I (FRI) and BL Lacs. It is understood that the fundamental connection between these types of AGN and Galactic BH-XRBs is simply their sub-Eddington accretion rates, and the presence of radio-emitting jets. All sources in the FP display something resembling a power law spectrum in the X-ray band, and a flat-to-inverted radio spectrum. Thus, by invoking this scaling relation, and assuming that an isolated population of low-luminosity accreting BHs will display similar spectral properties, one can adopt the FP relation to scale their X-ray and radio luminosities with BH mass. We therefore adopt the following scaling relation, determined in one of the most recent such FP studies \citep{Plotkin:2012}: 
\begin{equation}
\begin{split}
\log \left({\frac{L_{\rm X}}{\rm erg \, cm^{-2} s^{-1}}}\right) = (1.45 \pm 0.04) \log \left({\frac{L_{\rm R}}{\rm erg \, cm^{-2} s^{-1}}}\right)\\ - (0.88 \pm 0.06) \log  \left( \frac{M_\mathrm{BH}}{M_\odot} \right) - 6.07 \pm 1.10,
\end{split}
\label{eq:FP}
\end{equation}
Where $L_X$ is the X-ray luminosity in the 2-10 keV band and $L_R$ is the radio luminosity at 5 GHz. Thus we have a simple prescription for both the X-ray luminosity, $L_{\rm X}$, and radio luminosity, $L_{\rm R}$, of an isolated accreting BH.

\vspace{1cm}

\section{Estimating the number of visible sources}
\label{sec:estimate}

Having described our prescription for the accretion rate and the associated non-thermal emission, we now turn at assessing the number of isolated astrophysical black holes that are potentially detectable by the current (and forthcoming) generation of X-ray (with particular focus on the hard X-ray band) and radio experiments. In this section we summarize the main points of our methodology.
Our main observable is the number of sources associated to a radiation flux above detection threshold, i.e. that satisfy $ \phi > \phi^*$, where $ \phi$ is the flux at Earth defined as:
$
     \phi= L/ (4 \pi r^2)
$
with r the distance of the source from Earth, and $ \phi^*$ is the threshold value.  

Previous studies (\citealt{Fender:2013, Tsuna:2018oqt, Tsuna:2019kny}) obtained their estimates through Montecarlo simulations of a BH population. Here we adopt instead a semi-analytical approach which allows us to perform a comprehensive parametric study associated to the physical model of accretion physics described above.

Here we describe the general prescription for computing the number of visible sources, which we will apply with some variation to the two region of interest for this work: the local region (section \ref{sec:searchlocal}) and the central molecular zone (CMZ, section \ref{sec:searchcmz}).
We obtain the probability for a BH to emit above threshold by integrating the joint p.d.f. of the relevant randomly distributed variables over the volume defined by the condition $ \phi > \phi^*$ . 

Regarding the BH population, the random variables entering the flux expression are: speed, mass and distance from Earth. We neglect any possible correlation between these variables. In addition, we also treat as a random variable the density of the insterstellar medium at each BH location. In both analysis we present below, we make the assumption that the gas density and BH position are not correlated. 
Therefore, we obtain the expected number of luminous sources $N_{\rm sources}$ by computing the following integral:
\begin{equation}
\label{eq:Nsources}
\begin{split}
    &N^{\rm sources}  ( \phi^*, \{ p_i\}, \{ q_i\}) \,\,=\\
    N^{\rm tot} & \int_{\phi ({\rm v}_{\rm BH}, M, d , \{ p_i\}) > \phi^*} P({\rm v}_{\rm BH})P(M)P(r)P(n) \,d{\rm v}_{\rm BH}\,dM\,dr \,dn
\end{split}    
\end{equation}
Here $P({\rm v}_{\rm BH})$, $P(M)$, $P(r)$ and $P(n)$ are the normalized p.d.fs describing respectively the BH speed, mass, distance from Earth and the interstellar medium density at its location. With $\{p_i\}$ we indicate the free parameters entering the expression for the flux, and which we consider to be fixed for all BHs. These are: the sound speed of the neutral medium $c_s$, the sound speed in  the ionized region $c_s^{in}$, the fraction of bolometric luminosity $L_x/L$ and, in the case of BHL accretion, the suppression factor $\lambda$ (regarding the first of these, we remark that variations of the sound speed of the neutral medium $c_s$ have a negligible impact on the flux magnitude). Finally, the parameters that define the p.d.f.s for the the random variables, which we indicate with $\{q_i\}$, should also be regarded as free parameters of the model. We discuss them in the next section.

 \section{Characterizing the Black Hole population}
 \label{sec:bhpopulation}
 
 \noindent
{\bf Mass.} 
We assume BH masses to follow a normal distribution of mean $\mu_{\rm mass} = 7.8 \, M_\odot$ and $\sigma_{\rm mass} = 1.2 \, M_\odot$. This distribution was obtained from the study of X-ray binaries by \cite{_zel_2010} and confirmed in an independent analysis by \cite{Farr_2011}. It has also been employed by previous studies on this topic such as the works of \cite{Fender:2013} and \cite{Tsuna:2018oqt}. 
 
 \noindent
 {\bf Speed.}
 The BH velocity is given by the combination of two components: the velocity of the progenitor star, ${\rm v}_{\rm star}$, and the kick the BH receives at birth due to the supernova explosion, ${\rm v}_{\rm kick}$. 
 As for the former, we are concerned with the velocities relative to the molecular gas, so we ignore the rotational component along the Galactic disk and consider exclusively the velocity dispersion $\sigma_{D}$. This is well measured for both nearby stars and stars in the Galactic centre. 

 On the other hand, little is known about the magnitude of the natal BH kicks. A study of pulsars proper motions by \cite{Hobbs_2005} concluded that neutron star kicks obey a Maxwell Boltzmann distribution with $\sigma = 265$ km/s. Re-scaled with the average masses \citep{Fender:2013}, this gives $\sigma_{\rm kick} \approx 50 $ km/s, corresponding to a mean of $\mu_{\rm kick} \approx 75$ km/s. We consider here as reference values $\mu_{\rm kick} =50$ km/s ({\it"low kick"}) and $\mu_{\rm kick} =100$ km/s ({\it"high kick"}). 
 
The BH speed, resulting from the combination of these two independent components, is then distributed following a Maxwell Boltzmann of mean:
$
     \mu_{\rm BH}=\sqrt{\mu_{\rm star}^2+\mu_{\rm kick}^2},
 $.
 Hence, the uncertainties in both the kick and the progenitor star's velocity dispersion are enclosed in only one parameter, $\mu_{\rm BH}$.

 \noindent
 {\bf Distance from Earth.}
 We will assume the BHs to be distributed uniformly in both regions under analysis and derive the distance distribution accordingly.
 
  \noindent
  {\bf Density of the interstellar medium. }
  The interstellar medium is usually described as composed by three different components: the ionized component, the neutral gas, and the molecular clouds. In the PR13 picture the probability of obtaining large enough fluxes from a BH found in the first two of these components turns out to be negligible.
  We will therefore consider only densities associated to molecular clouds.

\section{Searching for black holes in the local region}
\label{sec:searchlocal}

We start by applying our model to the study of the innermost 250 pc around Earth, following the work of \cite{Fender:2013}. In this study, the authors applied the Bondi formula to model accretion and estimated the number of visible sources through Montecarlo simulations. They considered a few combinations of the values of the mean BH speed $\mu_{\rm BH}$ and of the suppression factor $\lambda$, obtaining a bound on the combination of these parameters. Here we reproduce the same setup to obtain a prediction for the number of detectable X-ray sources, verifying  the result for the BHL model and comparing it to the predictions of the PR13 model. 

\subsection{Our setup}

Regarding the BH population, we assume $N^{\rm local}=3.5 \times 10^4$ objects are present in the region between $70\, \text{pc} < \text{d} <250\, \text{pc}$ \citep{Fender:2013}. We assume a uniform spatial distribution, a normal mass distribution and a MB velocity distribution of average $\mu_{\rm BH}$, as described in the previous section. 
As for the interstellar medium, we assume two types of molecular clouds to be present, each type with uniform density. These are: a) warm clouds, of number density $10^2$~cm$^{-3}$, sound speed 0.9 km/s, filling factor $5 \times 10^{-2}$, and b) cold clouds, of number density $10^3$~cm$^{-3}$, sound speed 0.6 km/s and filling factor $5 \times 10^{-3}$ \citep{Fender:2013}. This combination results in an average density $n_{\rm avg}= 10 \, cm^{-3}$. The integral in Eq. \ref{eq:Nsources} becomes:
\begin{equation}
N^{\rm sources}  = \sum_{\rm warm, cold} N^{\rm clouds} \int_{\phi > \phi^*} P({\rm v}_{\rm BH})P(M)P(r) \,d{\rm v}_{\rm BH}\,dM\,dr  \end{equation}
Here $N_{\rm clouds} $ indicates the number of BHs we expect to find in each type of cloud: 
\begin{equation}
 N^{\rm clouds}=f_{\rm clouds} \, N^{\rm local},
\end{equation}
where $f_{\rm clouds} $ is the associated filling factor. We obtain $N_{\rm clouds} \approx 175$ in the cold clouds and $N_{\rm clouds} \approx 1750$ in the warm ones.

Finally, the bolometric fraction of luminosity in the X-ray band is set to $0.3$.

\subsection{Results}

\begin{figure}
  \centering
  \includegraphics[width=\linewidth]{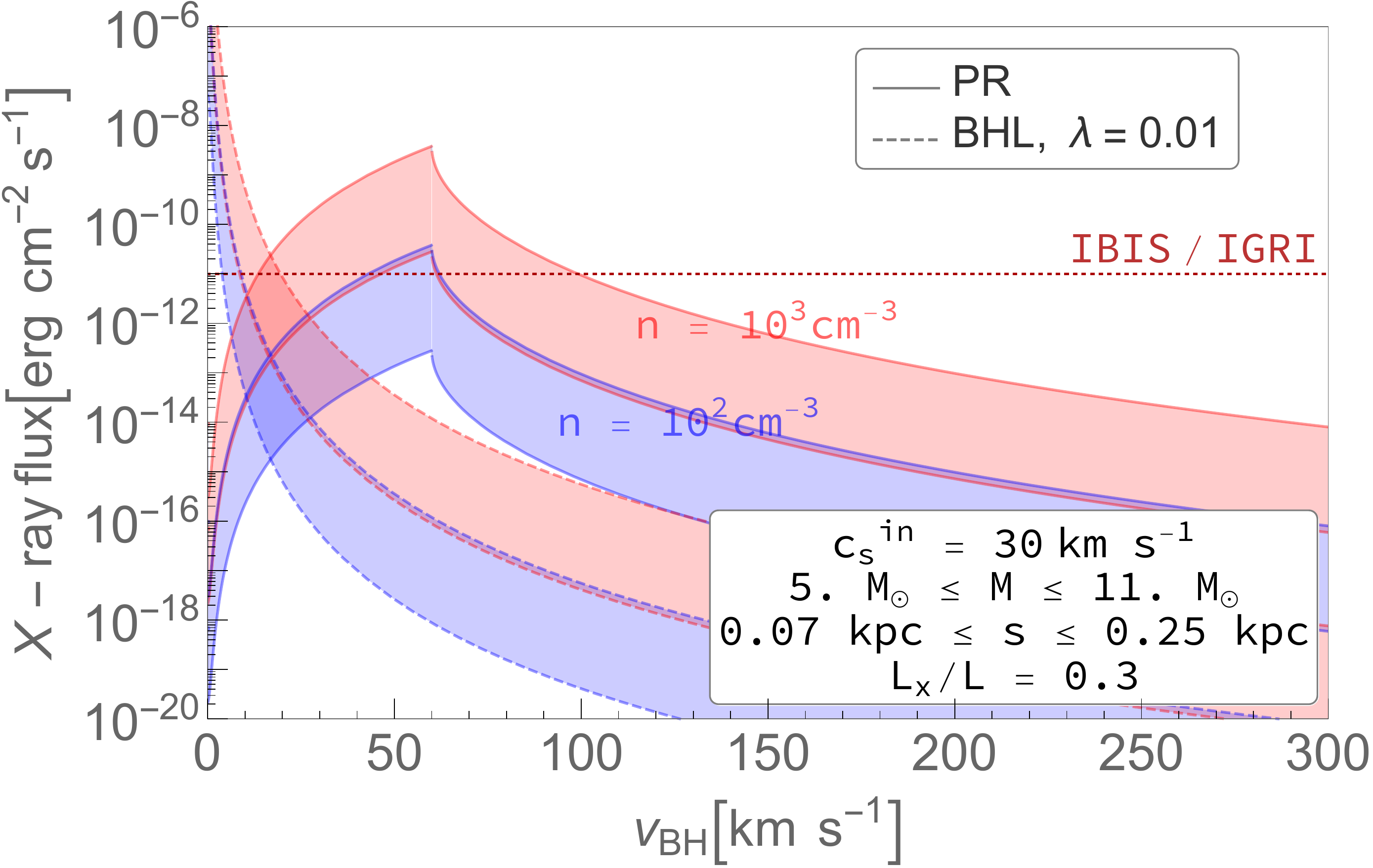}
\caption{\it {\bf X-ray flux from BHs in the local region.} We show the X-ray flux associated to a BH accreting from a molecular cloud in the local region, as a function of the relative velocity with respect to the gas cloud. The colored bands describe the two types of clouds considered and their width corresponds to a range of BH masses and distances. We show the fluxes for the Park-Ricotti and the suppressed ($\lambda=0.01$) Bondi-Hoyle-Littleton models. As a reference, we show the detection threshold associated to the IBIS/IGRI survey \citep{Bird_2009}.}
\label{XFluxLocal}
\end{figure}

\begin{figure}
\centering
  \includegraphics[width=\linewidth]{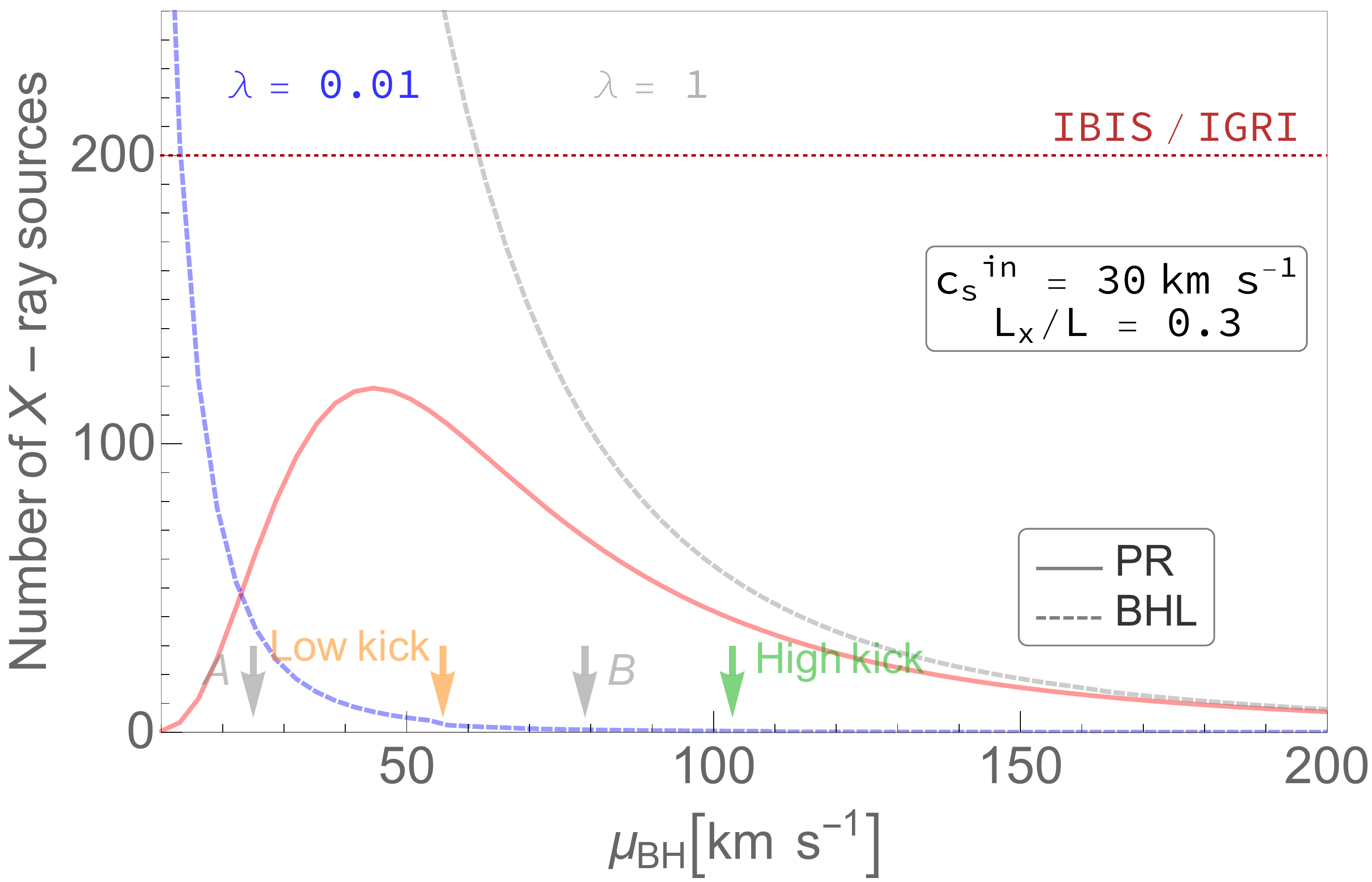}
\caption{\it {\bf Total number of X-ray sources in the local region (full sky, d < 250 pc).} We show the expected number of BH observed in the X-ray as a function of the average BH speed according to a) the PR13 model b) the BHL model with $\lambda$=1) the BHL model with $\lambda$=0.01. 
}
\label{fig:NLocal}  
\end{figure}

Let us first consider the X-ray flux associated to a generic isolated BH located in the local region and accreting gas from a molecular cloud as a function of the BH speed. In figure \ref{XFluxLocal} we show this flux for both the PR13 accretion scenario and the conventional modeling based on the BHL formalism, the latter suppressed by a factor $\lambda =0.01$. The two colored bands correspond to the two densities considered for molecular clouds. The width of the band is obtained considering distances between 70 and 250 pc and masses between 5 and 11 solar masses.  We show for reference the detection threshold associated to the 4th INTEGRAL IBIS/ISGRI soft gamma-ray
survey catalogue (17–100 keV; \citealt{Bird_2009}) 
, which is approximately $10^{-11}$ erg/cm$^2$/s. 
We want to emphasize once again the distinct behaviour associated to the two accretion scenarios.
In the Bondi picture, the slower sources are very luminous. This is particularly relevant in this context, since velocities in the local region are expected to be on the low end of the range shown in this figure: the unsuppressed BHL scenario predicts a huge number of sources (see figure \ref{fig:NLocal}) and can easily overshoot the bound. As a consequence of the introduction of a suppression factor, only very slow sources are expected to be visible. This is no longer true when the PR13 scenario is considered, since it naturally predicts a flux suppression for the slower sources. Instead, the population of BHs emerging above threshold and showing up in the X-ray sky will present relatively high speeds (around $50$ km/s). Furthermore, we can notice how, based on the Bondi picture, we expect to detect BHs in both types of clouds. According to the PR13 model, however, only BHs located in the denser clouds can be detected.

Following the procedure described in the previous section, we can now integrate the distributions that characterize the BH population to obtain the number of sources visible in the X-ray sky with the IBIS/IGRI survey. 
In figure \ref{fig:NLocal} we show the dependence of number of sources on the average speed $\mu_{\rm BH}$. This prediction is obtained for three accretion scenarios: {\it a)} the PR13 model; {\it b)} the BHL model with suppression factor $\lambda$= 0.01 c) the BHL model with perfect efficiency $\lambda$= 1. 
We indicate with orange and green arrows the reference values given by the {\it "low kick"} the {\it "high kick"} scenarios, considering a velocity dispersion of the progenitor stars of $15$ km/s (corresponding to an average speed of $\approx 25$ km/s). For comparison, the average speeds corresponding to the {\it "no kick"} and {\it"high kick"} ( $\mu_{\rm  kick}$ = 75 km/s) scenarios considered in \cite{Fender:2013} are also shown and indicated with A and B, respectively.

Our results can be summarized as follows:
\begin{itemize}
\item As far as the BHL scenario is concerned, our findings are in agreement with the ones obtained by the results reported in \cite{Fender:2013}: i.e., a few tens of visible sources in scenario A (no kick, $\lambda = 0.01$ ). Larger values of $\lambda$ are excluded unless very high kicks are present. This is, once again, due to the rapid increase of the BHL rate at low speeds. 
\item In the PR13 scenario, the introduction of a suppression factor is not necessary. While being compatible with the experimental bound, our model nevertheless predicts a significant number of bright sources. This prediction corresponds to a population of  $\sim40\mbox{--}100$ accreting isolated black holes in the existing catalogues, taking as reference the {\it "no kick"} and {\it"high kick"} scenarios . We will discuss in more detail the consequences of this result in the next Sections, and compare this prediction with the one associated to the Central Molecular Zone.
\end{itemize}

\section{Searching for black holes in the Central molecular zone}
\label{sec:searchcmz}

In this section we turn our attention to the inner part of the Galactic bulge. 
This region is a particularly promising target for BH searches because of the presence of a large reservoir of molecular gas called {\it central molecular zone} (CMZ), a cloud complex with asymmetric shape that extends up to $\simeq 200$ pc away from the Galactic centre. 
The CMZ has been the object of extensive multi-wavelength observational campaigns over the years, aimed at characterizing its structure and physical properties, including star formation rate (see for instance \cite{2019MNRAS.484.5734K} and references therein for a recent analysis). 
The abundance of gas in the form of giant molecular clouds, and the location close to the gravity centre of the Galaxy clearly makes it the ideal target for our study.

\subsection{Our setup}
\label{sec:CMZsetup}

To model the CMZ, we employ a simplified version of the model by \cite{Ferriere2007}. We describe it as  a cylindrical region of half height 15 pc and radius 160 pc. 
The number of BHs contained in this region is of course very uncertain, but we can make a naive order-of-magnitude estimate, as follows.
We assume the BHs to be generated following the distribution of stars in the Galaxy. Here we are interested in the bulge component, which accounts for around $15 \%$ of the total and can be modelled  with a spherical exponential with scale radius $R_{\rm bulge} = 120 $ pc \citep{Sofue_2013}.  Integrating the bulge spherical exponential distribution over the CMZ volume gives us the fraction of ABHs born in this region: around $2 \%$ of the bulge component. Assuming a total of $10^8$ BHs present in the whole Galaxy, this corresponds to  $3 \times 10^5$ BHs. We must however take into account that, due to the large initial natal kicks, the initial spatial distribution is modified \citep{Tsuna:2018oqt}. We performed a simulation of the evolution in the Galactic potential (\cite{Irrgang_2013}, model II)  of 1000 BHs. These are initially distributed uniformly in the CMZ region with an average speed of 130 km/s and given an average natal kick of 75 km/s. The simulation shows that only $\approx 25 \%$ of the BHs remain in the region: we observe in particular a spreading in the direction perpendicular to the Galactic plane. On the other hand, we do expect some BHs born outside of the CMZ to enter the region under the effect of the gravitational potential. However, these objects cross the CMZ close to the periastron of their orbit, with very large velocities and hence suppressed accretion rates. We can therefore neglect the latter effect. However the former is significant and we take it into account. We thus update our naive estimate to $ N ^{ \rm CMZ} \approx 7.5 \times 10^4$.

The molecular clouds in this region are known to be denser than the Galactic average.
We assume an average number density of molecular hydrogen of $n_{\rm avg} = 150\, \rm cm^{-3}$ \citep{Ferriere2007} and consider two types of clouds: warm, less dense clouds, and cold, denser ones. The warm clouds are of secondary importance given the large distance from Earth (see figure \ref{XFluxCMZ}). We set their density to $n= 10^{2.5} \, {\rm cm }^{-3}$ and consider a filling factor of 14\% \citep{Ferriere2007}. Regarding the cold clouds, we choose to adopt a more refined modelling based on a power law density distribution between $n_{\rm min}$ and  $n_{\rm max}$ :

\begin{equation}
    P(n) \propto n^{- \beta}
\end{equation}

Based on \cite{Ferriere2007}, we set $n_{\rm min}=10^{3.5} \, {\rm cm }^{-3}$ and  $n_{\rm max}=10^5 \, {\rm cm }^{-3}$, while we treat $\beta$ as a free parameter. The filling factor for the cold clouds is then obtained by requiring  $n_{\rm avg} = 150\, \rm cm^{-3}$. For a reference value of  $\beta =2.4$ \citep{HeRG:2019,HeRG:2020} we obtain a filling factor of $\approx 1 \%$ and an average cloud density of $\approx 8 \times 10^3 $.

The temperature of the clouds enters the accretion rate through the sound speed $c_s$. As we mentioned before, this variable has little impact on the predictions and we set it to 1 km/s.

Regarding the speed distribution, the average speed $\mu_{\rm BH}$ is treated as a free parameter. Nevertheless, we can make some estimates of reasonable values. The most recent observations suggest a velocity dispersion of the central bulge stars of average  $\mu_{\rm bulge} \approx 130$ km/s (\citealt{Sanders_2019}, \citealt{2018}). This gives, for reference, $\mu_v$= 140 km/s in the {\it "low kick" } scenario and $\mu_v$= 160 km/s in the {\it "high kick" } scenario.

Finally, we approximate the distance distribution with a delta peaked at 8.3 kpc and we assume the normal mass distribution discussed in Section \ref{sec:bhpopulation}.

The integral to be computed becomes:

\begin{equation}
\begin{split}
N^{\rm sources}  =  N^{\rm clouds}_{\rm warm} \int_{\phi > \phi^*} P({\rm v}_{\rm BH})P(M) \,d{\rm v}_{\rm BH}\,dM + \\
N^{\rm clouds}_{\rm cold} \int_{\phi > \phi^*} P({\rm v}_{\rm BH})P(M)P(n) \,d{\rm v}_{\rm BH}\,dM\,dn
\end{split}
\end{equation}

The expected number of black holes in the clouds is obtained as:
\begin{equation}
 N^{\rm clouds}=f_{\rm clouds} \, N^{\rm CMZ},
\end{equation}
where $f_{\rm clouds} $ is the associated filling factor.

 In summary, the set of relevant parameters that play a key role in our calculation are: 1) the average BH speed, $\mu_{\rm BH}$  2) the slope of the power law density distribution, $\beta$ 3) the sound speed within the ionized region around the BH, ${c_s}^{\rm in}$; 4) the fraction of bolometric luminosity irradiated in the hard X-ray band $L_{\rm x}/L$;  5) the number of black holes in the CMZ, $N^{\rm CMZ}$.

\subsection{Results}

\begin{figure}
  \centering
  \includegraphics[width=\linewidth]{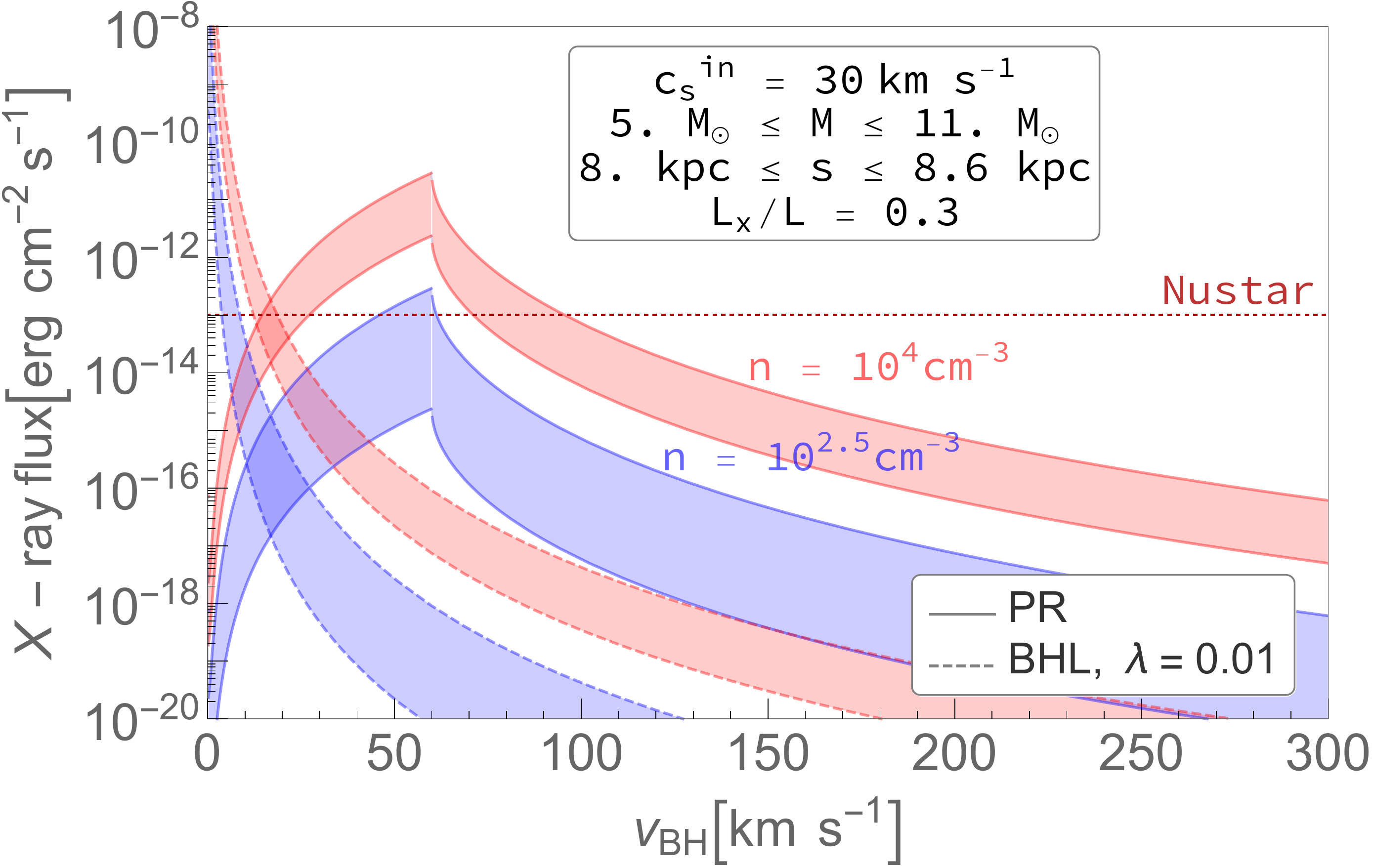}
\caption{\it {\bf X-ray flux from BHs in the CMZ.} We show the X-ray flux associated to a BH orbiting in the Central Molecular Zone molecular cloud complex, for different values of the relative velocity with respect to the gas cloud, and the gas density in the cloud. We consider the Park-Ricotti and the Bondi-Hoyle-Littleton formalism. As a reference, we show the detection threshold associated to the NuSTAR Galactic Center survey \citep{Hong:2016}.}
\label{XFluxCMZ}
\end{figure}

We start by considering the X-ray flux associated to a generic isolated BH located in the Central Molecular Zone and accreting gas from a molecular cloud as a function of the speed and of the molecular cloud density. We show this observable in figure \ref{XFluxCMZ} for the two accretion scenarios discussed above. As a reference, we show the detection threshold associated to the NuSTAR Galactic Center survey in the 3-40 keV band \cite{Hong:2016}. The same trends highlighted in Sec. 5.2 can be noticed in this plot: In particular, we point out that, if the cloud density is high enough, there is a wide range of BH speed associated to emission above threshold within the PR13 formalism. On the other hand, no sources are expected to be detected in the lower density clouds.

\begin{figure}
\begin{subfigure}
  \centering
  \includegraphics[width=\linewidth]{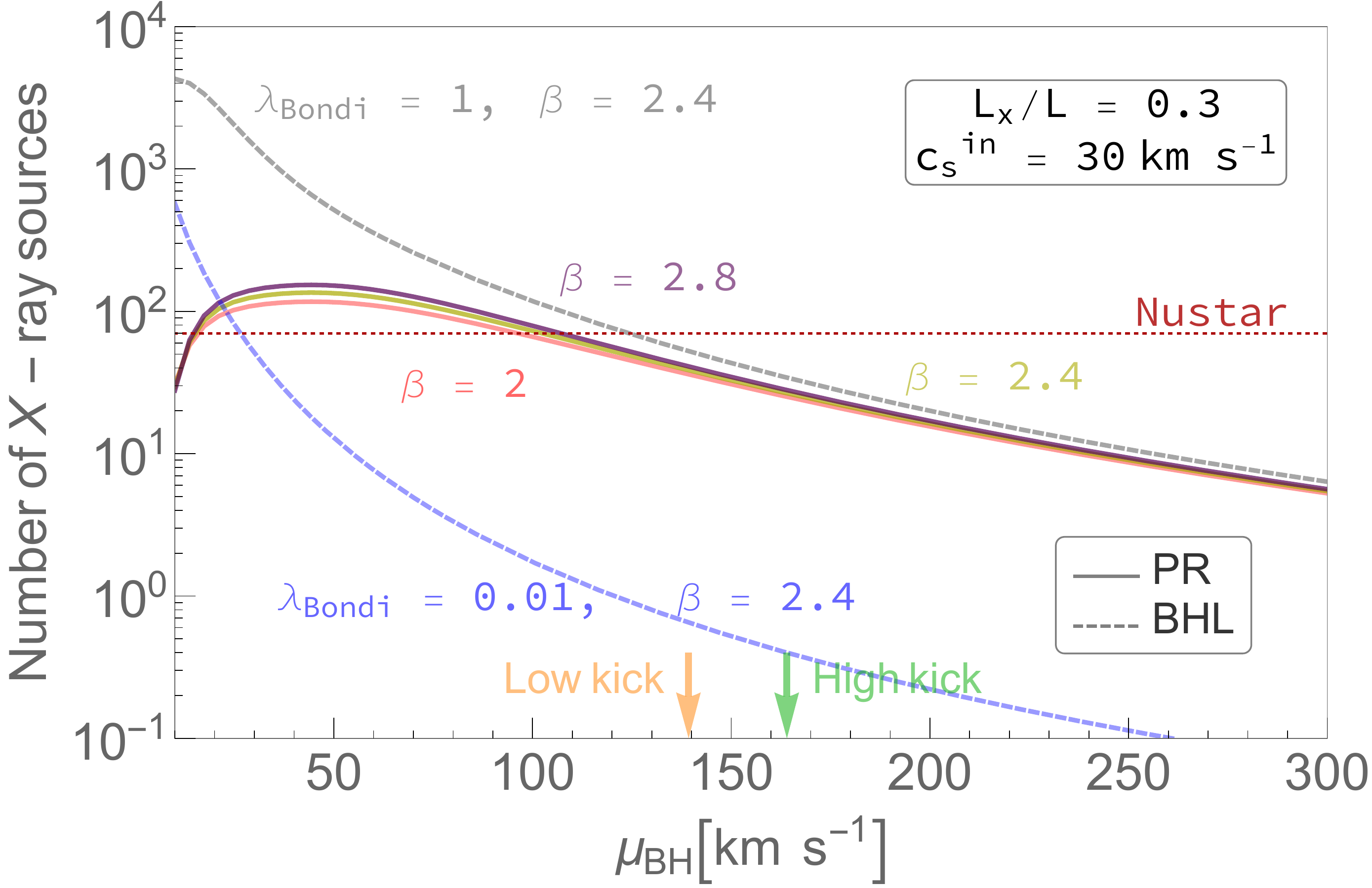}
\end{subfigure}
\begin{subfigure}
  \centering
  \includegraphics[width=\linewidth]{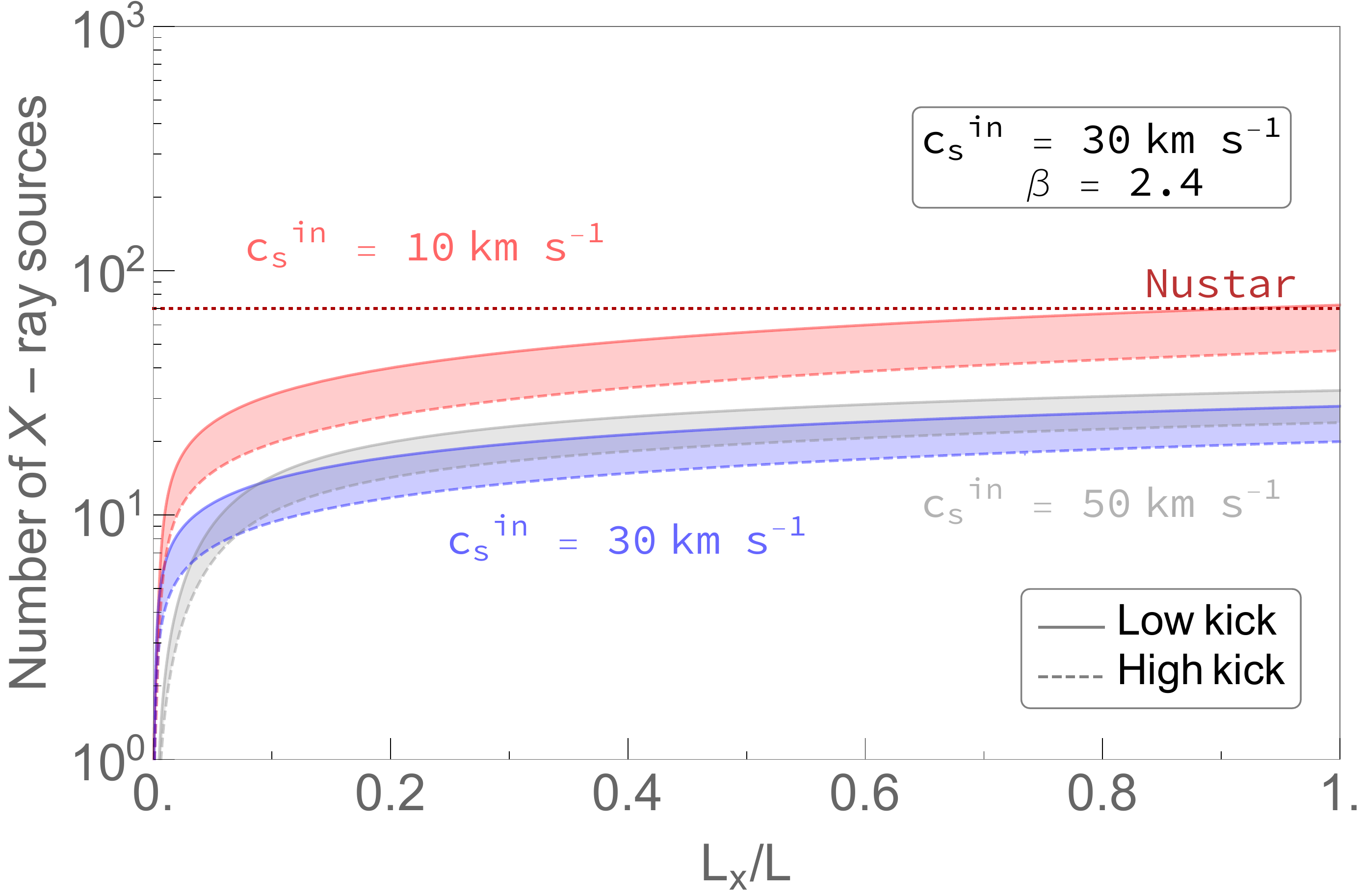}
\end{subfigure}
\caption{\it {\bf Parametric study of the number of X-ray sources in the CMZ}. Panel (a): Number of X-ray sources as a function of the average speed. Panel (b): Number of X-ray sources as a function of the bolometric luminosity that is radiated in the X-ray band of interest;  We compare this with the number of sources detected by NuSTAR in the CMZ region \citep{Hong:2016}.}
\label{NCMZ}
\end{figure}

With the X-ray flux for a CMZ source at hand, we apply the procedure described in the previous Sections, and compute the number of X-ray sources associated to accreting BHs in this region.
The results are shown in figure \ref{NCMZ}. In this panel we show the impact of the different free parameters on this key observable.
In particular, we show in panel (a) the number of X-ray sources in the CMZ region as a function of the average speed of the BH population adopting both the BHL and the PR13 accretion models. We show the result for different choices of the sound speed in the ionized region, and for different choices of the $\lambda$ parameter in the BHL case. We can notice how, in this setting, the PR13 scenario predicts as many sources as the $\lambda =1$ BHL, due to the fact that high speeds are prevalent among the BH population of the CMZ. However, we have seen that the the un-suppressed BHL scenario is excluded by previous studies on nearby compact objects and complementary studies focused on AGN populations as mentioned in Sec. \ref{sec:BHL}. 
The comparison should therefore be made between the predictions of the PR13 model and the suppressed BHL model, also shown in figure for $\lambda =0.01$. Then, the PR13 model predicts a significantly greater number than those obtained assuming BHL accretion. (See also the work of \cite{Tsuna:2018oqt}, in which $\mathcal{O}$(1) sources were predicted in the Galactic centre using BHL accretion).  
In panel (b) we focus on the dependence of the same observable with respect to the fraction of the bolometric luminosity that is radiated in the X-ray band of interest; in panel (c) we consider the differential contribution of the clumps of different density, again for different reference values of the ionized sound speed in the BH vicinity $c_s^{in}$.

\begin{figure}
  \centering
  \includegraphics[width=\linewidth]{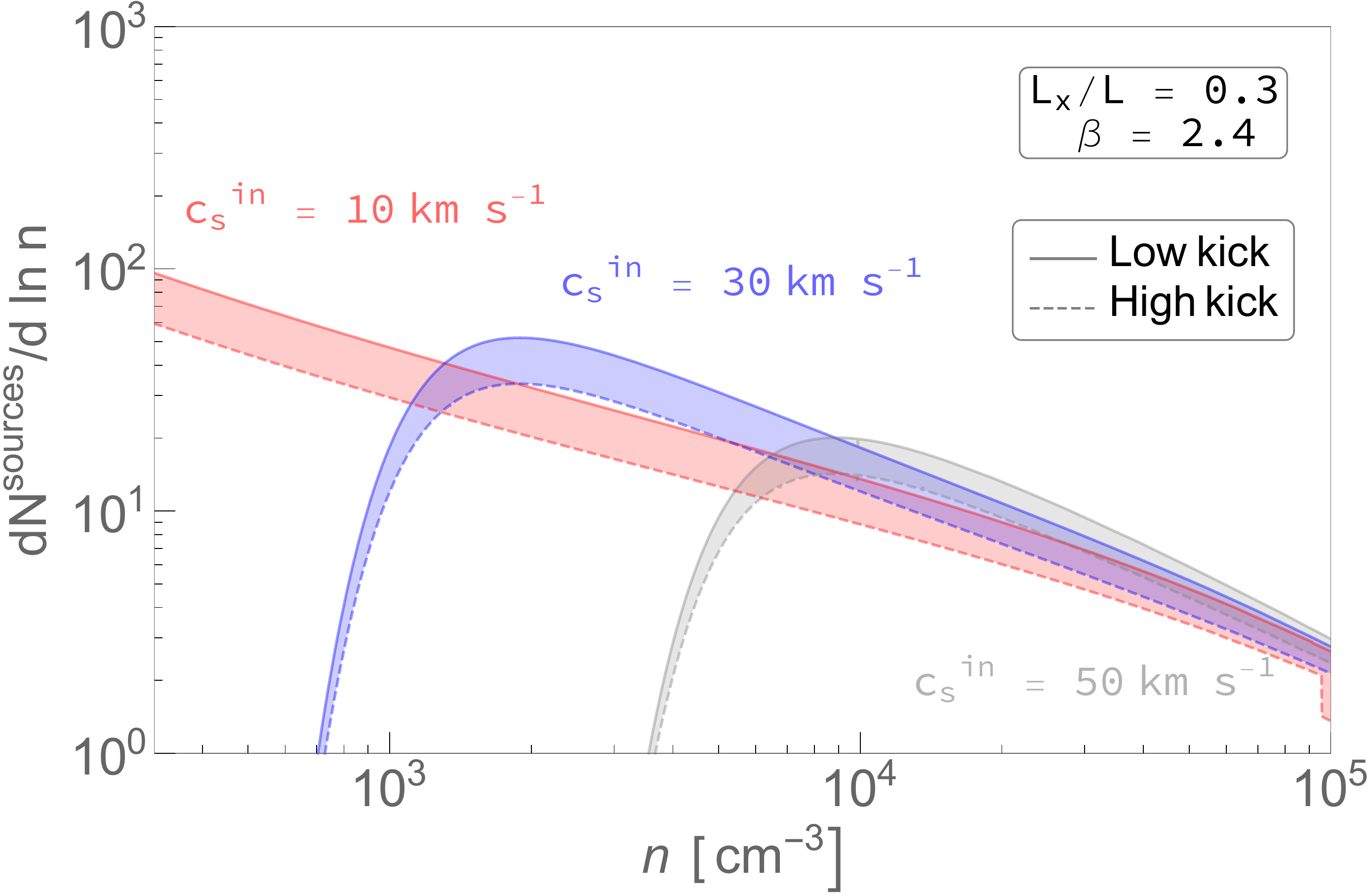}
  \caption{{ \bf Distribution of bright X-ray sources as a function of clump density.} We show the log-differential number of sources detected by NuSTAR as a function of the clump density. A power law distribution for the clump density is assumed.}
  \label{}
\end{figure}

We can conclude from the results visualized in figure \ref{NCMZ} that, despite the relevant uncertainties associated to the modelling of such a complex setup (and despite the existence of a threshold effect due to the peak in the accretion rate), the prediction of few tens of sources is solid with respect to these uncertainties. In particular, we notice from panel (a) that a number of bright X-ray sources comprised between $10$ and $20$ is expected for our reference value of $c_s^{in}$ and for $\mu_{BH} = 150$ km/s. As far as the speed distribution is concerned, we remark that the two reference values we have chosen to bracket the uncertainty both lead to similar predictions, while large deviations from this range would require to assume unrealistically high speeds. The same line of thought applies to the slope associated to the clump density distribution.
From panel (b) we also notice a lower number of sources may only arise if we were to assume either a very high temperature in the ionized region, or else a very low fraction of bolometric luminosity radiated in the X-ray band. We remark,however, that the prediction scales linearly with the expected number of BHs in the region, $N^{\rm CMZ}$, which carries a large uncertainty.

In summary, the PR13 model (together with our rather conservative assumptions regarding the X-ray emission) leads to the  {\it prediction of a significant number of X-ray sources in the CMZ region} associated to isolated BHs accreting from molecular clouds. The total number of sources is of the same order of magnitude as the one predicted in the local region and discussed in the previous Section. However, while the previous Section was dealing with a full-sky analysis, in this case we are focusing on a specific region of interest (ROI) with small angular extent, which is ideal for multi-wavelength observational campaigns.
We recall that NuSTAR has detected $70$ hard X-ray sources sources in this region interest. The nature of most of the sources in the catalogue in not precisely known, although a significant population of cataclysmic variables and X-ray binaries is expected. Our finding seems to suggest the need of a careful data analysis, in order to identify the possible presence of a relevant population of isolated accreting black holes in the already existing data. Following this line of thought, and in the prospect of a discovery, a multi-wavelength analysis is compelling, so we will turn our attention to the radio domain in the next Section.

\section{Multi-wavelength prospects for the Square Kilometer Array}
\label{sec:SKA}

\begin{figure}
  \centering
  \includegraphics[width=\linewidth]{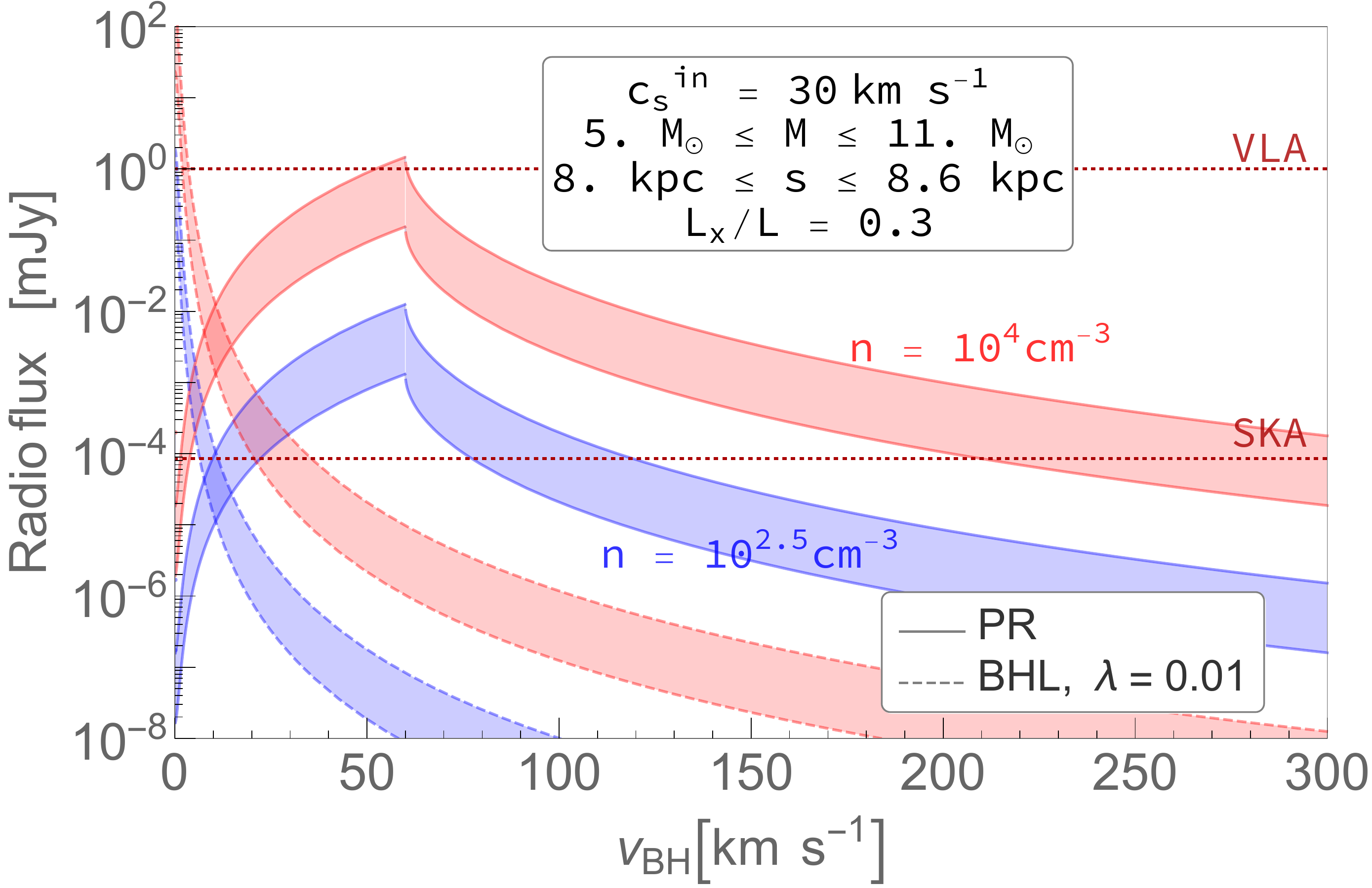}
\caption{\it {\bf Radio flux from BHs in the CMZ}: We show the radio flux for a generic BH in the CMZ. The present VLA sensitivity and the prospective sensitivity for SKA are shown. }
\label{fig:radioflux}
\end{figure}

\begin{figure}
  \centering
  \includegraphics[width=\linewidth]{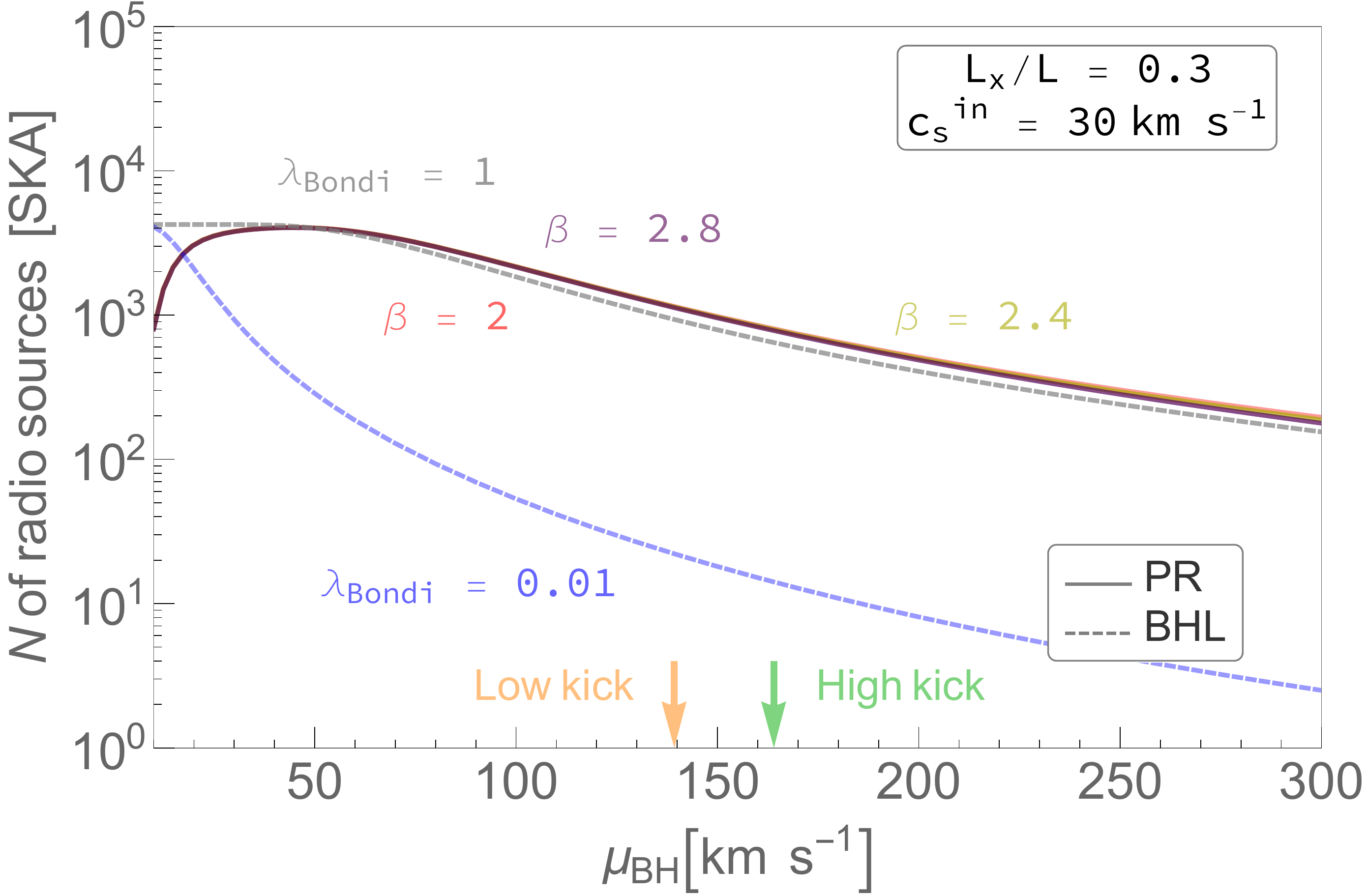}
\caption{\it {\bf Radio prediction (I)}: We show the number of radio sources expected from the CMZ as a function of the number density of gas in molecular clouds, for different choices of the relevant parameters, obtained without taking into account the information from the X-ray band. The combination of this prediction with the X-ray bound is presented in figure \ref{fig:SKA2}. 
}
\label{fig:SKA1}
\end{figure}

\begin{figure}
\centering
  \includegraphics[width=\linewidth]{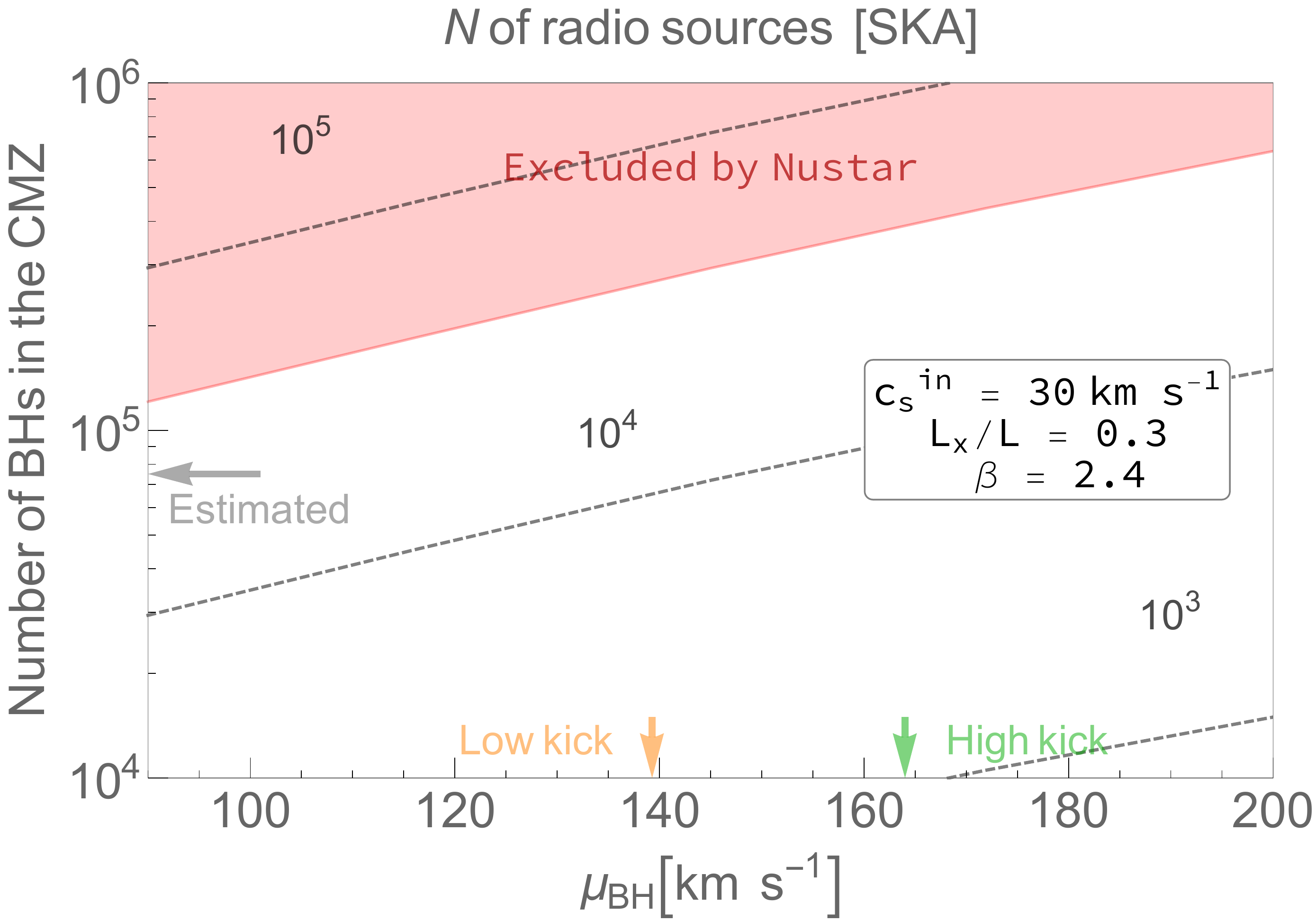}
\caption{\it {\bf Radio prediction (II)}: We show the contour lines for the number of sources potentially detectable by the Square Kilometer Array as a function of cloud density and total number of BHs in the CMZ clouds. The red region corresponds to the parameter space region excluded by NuSTAR observations. The grey arrow indicates the estimate of the number of BHs in the CMZ region used as reference in the previous section. 
}
\label{fig:SKA2}
\end{figure}

In the previous section we have shown that, in a wide portion of the parameter space associated to our problem, a large number of bright X-ray sources are expected in the Galactic Center region.
However, in order to pinpoint a source as an accreting black hole, a careful multi-wavelength study has to be performed. The GHz radio band is particularly interesting in this context. As in previous works, (\citealt{Gaggero:2016dpq,Manshanden:2018tze}) the empirical scaling between X-ray and radio flux (\citealt{Plotkin:2012}) discussed in detail in Section 3 (eq. \ref{eq:FP}) is employed to obtain the radio flux. 

A remarkable increase in the sensitivity is expected in the radio domain over the coming decade, thanks to the development of the Square Kilometer Array (SKA) project. This experiment has a huge potential towards shedding light on key problems of fundamental physics, cosmology and astrophysics \citep{Bull:2018lat}. Here we focus on the discovery potential of the population of astrophysical black holes in the CMZ at $1.4$ GHz band, and provide estimates of the number of potentially detectable sources by the SKA1-MID facility.
Assuming gain $G = 15$ K/Jy, receiver temperature $T_{\rm rx} =  25$ K, sky temperature towards the Galactic Center $T_{\rm sky} = 70$ K, and bandwidth $\delta \nu = 770$ MHz (as in \citealt{2016ApJ...827..143C}), we obtain an instrumental detection sensibility of $2.7 \mu$Jy for one-hour exposure. 
In the following, we assume an optimistic $1000$ h exposure time and consistently adopt a $85$ nJy as potential detection threshold.  

In figure \ref{fig:radioflux} we show the radio flux associated to a generic BH in the CMZ cloud, and compare it with the prospective SKA sensitivity and with the threshold of the VLA catalogue \citep{2008ApJS..174..481L}. Focusing on the PR13 scenario, we can conclude from this plot that, while we do would not expect to detect any isolated BH give the VLA sensitivity , SKA would be able to unveil a huge population of isolated BH in the Galactic centre. In fact, most of the BHs accreting from the cold clouds would emit above the SKA threshold.

We show our prediction for the number of radio sources detectable with SKA in figure \ref{fig:SKA1} and figure \ref{fig:SKA2},  obtained with the same setup described in section \ref{sec:CMZsetup}.
In particular, in figure \ref{fig:SKA1} we show the number of radio sources detectable by SKA as a function of the BH speed, for different choices of the parameters discussed in the previous Section. We can notice how our model leads to predict thousands of visible sources the two reference values we have chosen to bracket the uncertainty on the speed distribution.

Let us now widen our perspective and promote $N^{\rm CMZ}$, the total number of BHs in the CMZ to a free parameter.  In figure \ref{fig:SKA2} we show the number of bright radio sources in the ($\mu_{BH}$, $N^{\rm CMZ}$) space.
In this parameter space, we may use the comparison with NuSTAR observations discussed in the previous Section to obtain a bound: in particular, we can identify an upper limit on the number of black holes present in the CMZ, by requiring not to overshoot the number of sources observed by that experiment. 
We find that, in the allowed portion of parameter space, a very large number of radio point sources will be detectable by SKA. In summary, {\it the radio searches for isolated BHs in the CMZ constitute a promising science case for this experiment.}

\section{Discussion}
\label{sec:discussion}

{\bf Comments on the parameter space.}
We have concluded that our predictions for the CMZ are stable in a very wide region of the parameter space of the model. However, we should note that we made a number of assumptions regarding the distribution of black holes in the CMZ. On the one hand, as we already discussed, our estimate of the total number of BH in the CMZ is obtained with a simple model for the initial distribution and subsequent orbit evolution. Furthermore, we have assumed that both the BHs and the clouds are uniformly distributed in the CMZ. The presence of a positive or negative correlation between the two distributions could in principle impact the results. We further assumed a specific model for the cloud density distribution.
These simplifications eventually affect the number of BHs in the clouds, and therefore the final predictions on the number of sources, which scale linearly with this quantity.

{\bf The accretion scenario.}
 We have shown in Section \ref{sec:searchlocal} that, based on local observations, the PR13 accretion model does not require the introduction of a suppression factor $\lambda$. However, since the PR13 model relies on the BHL model to describe accretion within the ionized region (see Sec. \ref{sec:accretion}), one may wonder whether the BHL suppression factor should be introduced at that stage. With respect to this, we should take into account that this suppression factor is usually associated to a very strong outflow of matter from within the Bondi radius, associated to a fraction of $1-\lambda$ of the total matter initially accreted. Such a large outflow is not observed in the simulations of \cite{Park:2012cr}, where most matter crossing the Bondi radius within the ionized region ultimately reaches the BH. We conclude from this that introducing a suppression factor in the expression would not be physically justified.
We remark that the simulations considered here represent the state of the art in the field. However, some physical ingredients are still not captured: we mention in particular the role of magnetic fields. Future works will assess the potential impact of a full magneto-hydrodynamical treatment in this context.

{\bf The emission mechanism and spectrum.}
In Sec.~\ref{sec:CMZsetup} we also assume that $\sim30\%$ of the total bolometric luminosity of all our isolated BHs falls within the observable X-ray band---$L_{\rm X}=0.3~L_{\rm bol}$. This assumption was based on previous estimates, for example those presented by \cite{Fender:2013}. However, it is worth assessing the accuracy and precision of this assumption. We assumed a flat radio spectrum ($L_{\nu}\propto\nu^0$), and then an additional inverted power law in the X-ray band. The assumption that $30\%$ of the observed bolometric luminosity falls in the X-ray band depends on the spectral index of this additional power law, $\alpha$, the high-energy cut off of the power law, $E_{\rm cut}$, the absorbing hydrogen column density along the line of sight, $N_{\rm H}$, and the observing energy band (which we conservatively adopted as $3\mbox{--}40$~keV for the {\it NuSTAR} X-ray telescope). Inefficiently radiating Galactic BH-XRBs are dominated by power law emission with a typical range for the spectral index given by $0.6 \le\alpha\le 1.0$ (or a photon index of $1.6 \le\Gamma\le 2$ in common X-ray astronomy parlance). However, the power law spectra of Galactic BH-XRBs can become softer in intermediate spectral states, or very quiescent states, so we can extend this out to an upper bound of $\alpha\le 1.4$ or $\Gamma\le 2.4$. Adopting a rough estimate of the jet break frequency of $\sim10^{14}$~Hz (infrared frequencies; \citealt{Russell2013}), a typical high-energy cut off in the X-ray of $100$~keV, and a high hydrogen column density of $10^{23}~{\rm cm^{-2}}$, we calculate that $L_{\rm X}\sim(0.24\mbox{--}0.35)~L_{\rm bol}$. The corresponding unabsorbed range of fractional luminosities is then $0.27\mbox{--}0.46~L_{\rm bol}$. These predictions, however, depend strongly on the cutoff energy. For example, just reducing the cutoff energy from $100$~keV to $80$~keV can increase the fractional luminosity to $\sim60\%\mbox{--}70\%$. Therefore, our assumption that $L_{\rm X}\sim0.3~L_{\rm bol}$ is rather conservative, even if we assume strong absorption along the line of sight. It is also based upon a relatively hard power law spectrum in the X-ray band. Nevertheless, we have shown in Section \ref{sec:searchcmz}, that our predictions are solid with respect to variations of this parameter, even for values as large as $70 \%$.

For the radio flux calculations through the Fundamental Plane of Black Hole Activity (FP; \cite{Merloni:2003,F04,Plotkin:2012}) we had to assume the presence of a jet emitting in the GHz radio band, following \cite{Fender:2013, Gaggero:2016dpq, Manshanden:2018tze}. This crucial assumption is motivated by results of \cite{Fender:2001}, which established that all highly sub-Eddington accreting BHs in X-ray binary systems are accompanied by such radio emission. 

Further emission could come from other types of outflow that are less collimated or relativistic. Such alternative emission could constitute a lower limit on the expected radio emission, which would be of particular interest in the absence of a jet. \cite{Tsuna:2019kny} assumed a spherically symmetric outflow based on losing a fraction (1-$\lambda$) of the accreting matter before reaching the BH, where $\lambda$ is the suppression factor. Assuming a power-law radial dependence of the accretion rate, they used the escape velocity to conservatively estimate the power associated to such an outflow. It should be noted that this radial power-law assumption (with slope larger than $1$) implies that most matter would be lost close to the BH, which requires a powerful outflow. The shock from the collision of this outflow with the ISM could 
accelerate non-thermal electrons through diffusive shock acceleration and subsequently produce radio-synchrotron emission. The precise emission profile will be highly dependent on the fraction of energy going into the electrons and the magnetic field.

However, as highlighted above, such a relevant outflow is not observed in the simulations presented in \cite{Park:2012cr} and hence this scenario is hard to unify with the PR13 accretion model. 
Instead, it would be interesting to further investigate what type of emission could be associated to the shock on the boundary of the cometary region itself, which would likely require additional detailed simulations.

{\bf Connection with PBH searches.}
We now want to widen the perspective and mention that  the search for isolated astrophysical black holes (ABHs) can be considered a relevant problem {\it per se}, and also crucial as a background in the context of the quest for primordial black holes (PBHs) in the Galactic environment.
The potential of this observational channel in constraining the abundance of PBHs in the inner part of the Galaxy, with particular focus on the CMZ region, was first pointed out in \cite{Gaggero:2016dpq}, within the framework of the Bondi-Hoyle-Littleton formalism. 
In \cite{Manshanden:2018tze} the impact of the Park-Ricotti formalism was assessed in detail. The main result of that work is a strong upper bound on the abundance of PBHs (in the context of a non-clustered distribution of PBHs with monocromatic, log-normal and power-law mass functions). The upper bound was set at the level of $10^{-3}$ of the Dark Matter in the form of PBHs, which would correspond to roughly $\sim 10^8$ objects in the Milky Way for a $10$ M$_{\odot}$ reference mass. Hence, the astrophysical population (which is expected to include the same order of magnitude of compact objects) is expected to be an irriducible background that could play the role of an {\it astrophysical floor} associated to the quest for a subdominant population of PBHs of tens of solar masses.

{\bf Outlook.}
We have shown that a scenario based on the PR13 accretion model and our best estimate of the number of BHs in the CMZ naturally implies a {\it large number of detectable X-ray and radio sources}, for any reasonable choice of the other free parameters involved in the problem. 

This relevant result is highly suggestive that a clear identification of a population of isolated, accreting BHs in the Galactic Center region --- by means of the current or forthcoming generation of X-ray experiments --- may be around the corner.

\section{Summary}
\label{sec:summary}

In this paper we presented a comprehensive study of the constraints and prospects of detection of a population of isolated astrophysical black holes in the solar vicinity and near the Galactic Centre.
We have adopted the Park-Ricotti (PR13) accretion model that takes into account radiation feedback and is backed up by hydrodynamic numerical simulations, and compared the results obtained within this scenario with the ones obtained by exploiting the well-known Bondi-Hoyle-Littleton formalism.
We found that, within the PR13 model, a large number ($\sim40\mbox{--}100$) of bright X-ray sources associated to isolated accreting black holes is expected in the solar vicinity ($R < 250$ pc), and a significant fraction of the sources listed in current catalogues could be associated to such objects.
We performed an extended parametric study about the number of detectable sources in the vicinity of the Galactic Centre, with particular focus to the promising Central Molecular Zone region. In this region of interest we found that, despite the large uncertainty associated to the free parameters involved in the calculation, the prediction of $\mathcal{O}(10)$  X-ray sources above the detection threshold associated to the NuSTAR catalogue is solid for any reasonable choice of the parameters under scrutiny. In particular, a number of bright X-ray sources comprised between $10$ and $20$ is expected for our reference value of the ionized sound speed $c_s^{in}= 30$ km/s and an average black hole speed $\mu_{\rm BH}= 150$ km/s. 
A multi-wavelength analysis will be needed to clearly identify such a population. In this context, we pointed out promising prospects of detection for future generation experiments in the radio band, with particular reference to the Square Kilometre Array.

\section*{Acknowledgements}

The work of DG and FS has received financial support through the Postdoctoral Junior Leader Fellowship Programme from la Caixa Banking Foundation (grant n. LCF/BQ/LI18/11630014).

DG was also supported by the Spanish Agencia Estatal de Investigaci\'{o}n through the grants PGC2018-095161-B-I00, IFT Centro de Excelencia Severo Ochoa SEV-2016-0597, and Red Consolider MultiDark FPA2017-90566-REDC.

MR acknowledges support by NASA grant 80NSSC18K0527.

\section*{Data availability}

No new data were generated or analysed in support of this research.

\newpage

\clearpage
\bibliographystyle{mnras}
\bibliography{refs.bib}

\begin{thebibliography}{}
\makeatletter
\relax
\def\mn@urlcharsother{\let\do\@makeother \do\$\do\&\do\#\do\^\do\_\do\%\do\~}
\def\mn@doi{\begingroup\mn@urlcharsother \@ifnextchar [ {\mn@doi@}
  {\mn@doi@[]}}
\def\mn@doi@[#1]#2{\def\@tempa{#1}\ifx\@tempa\@empty \href
  {http://dx.doi.org/#2} {doi:#2}\else \href {http://dx.doi.org/#2} {#1}\fi
  \endgroup}
\def\mn@eprint#1#2{\mn@eprint@#1:#2::\@nil}
\def\mn@eprint@arXiv#1{\href {http://arxiv.org/abs/#1} {{\tt arXiv:#1}}}
\def\mn@eprint@dblp#1{\href {http://dblp.uni-trier.de/rec/bibtex/#1.xml}
  {dblp:#1}}
\def\mn@eprint@#1:#2:#3:#4\@nil{\def\@tempa {#1}\def\@tempb {#2}\def\@tempc
  {#3}\ifx \@tempc \@empty \let \@tempc \@tempb \let \@tempb \@tempa \fi \ifx
  \@tempb \@empty \def\@tempb {arXiv}\fi \@ifundefined
  {mn@eprint@\@tempb}{\@tempb:\@tempc}{\expandafter \expandafter \csname
  mn@eprint@\@tempb\endcsname \expandafter{\@tempc}}}

\bibitem[\protect\citeauthoryear{{Baganoff} et~al.,}{{Baganoff}
  et~al.}{2003}]{2003ApJ...591..891B}
{Baganoff} F.~K.,  et~al., 2003, \mn@doi [\apj] {10.1086/375145}, \href
  {https://ui.adsabs.harvard.edu/abs/2003ApJ...591..891B} {591, 891}

\bibitem[\protect\citeauthoryear{Bird et~al.,}{Bird et~al.}{2009}]{Bird_2009}
Bird A.~J.,  et~al., 2009, \mn@doi [The Astrophysical Journal Supplement
  Series] {10.1088/0067-0049/186/1/1}, 186, 1–9

\bibitem[\protect\citeauthoryear{{Blandford} \& {Begelman}}{{Blandford} \&
  {Begelman}}{1999}]{Blandford:1999}
{Blandford} R.~D.,  {Begelman} M.~C.,  1999, \mn@doi [MNRAS]
  {10.1046/j.1365-8711.1999.02358.x}, \href
  {http://adsabs.harvard.edu/abs/1999MNRAS.303L...1B} {303, L1}

\bibitem[\protect\citeauthoryear{{Bondi} \& {Hoyle}}{{Bondi} \&
  {Hoyle}}{1944}]{Bondi1944}
{Bondi} H.,  {Hoyle} F.,  1944, \mn@doi [MNRAS] {10.1093/mnras/104.5.273},
  \href {http://adsabs.harvard.edu/abs/1944MNRAS.104..273B} {104, 273}

\bibitem[\protect\citeauthoryear{Brown et~al.,}{Brown et~al.}{2018}]{2018}
Brown A. G.~A.,  et~al., 2018, \mn@doi [Astronomy & Astrophysics]
  {10.1051/0004-6361/201833051}, 616, A1

\bibitem[\protect\citeauthoryear{{Calore}, {Di Mauro}, {Donato}, {Hessels}  \&
  {Weniger}}{{Calore} et~al.}{2016}]{2016ApJ...827..143C}
{Calore} F.,  {Di Mauro} M.,  {Donato} F.,  {Hessels} J.~W.~T.,   {Weniger} C.,
   2016, \mn@doi [\apj] {10.3847/0004-637X/827/2/143}, \href
  {https://ui.adsabs.harvard.edu/abs/2016ApJ...827..143C} {827, 143}

\bibitem[\protect\citeauthoryear{{Caputo}, {de Vries}, {Patruno}  \& {Portegies
  Zwart}}{{Caputo} et~al.}{2017}]{2017MNRAS.468.4000C}
{Caputo} D.~P.,  {de Vries} N.,  {Patruno} A.,   {Portegies Zwart} S.,  2017,
  \mn@doi [\mnras] {10.1093/mnras/stw3336}, \href
  {https://ui.adsabs.harvard.edu/abs/2017MNRAS.468.4000C} {468, 4000}

\bibitem[\protect\citeauthoryear{{Corbel}, {Fender}, {Tzioumis}, {Nowak},
  {McIntyre}, {Durouchoux}  \& {Sood}}{{Corbel} et~al.}{2000}]{Corbel2000}
{Corbel} S.,  {Fender} R.~P.,  {Tzioumis} A.~K.,  {Nowak} M.,  {McIntyre} V.,
  {Durouchoux} P.,   {Sood} R.,  2000, \aap, \href
  {http://adsabs.harvard.edu/abs/2000A%26A...359..251C} {359, 251}

\bibitem[\protect\citeauthoryear{{Corbel}, {Nowak}, {Fender}, {Tzioumis}  \&
  {Markoff}}{{Corbel} et~al.}{2003}]{Corbel2003}
{Corbel} S.,  {Nowak} M.~A.,  {Fender} R.~P.,  {Tzioumis} A.~K.,   {Markoff}
  S.,  2003, \mn@doi [\aap] {10.1051/0004-6361:20030090}, \href
  {http://adsabs.harvard.edu/abs/2003A%26A...400.1007C} {400, 1007}

\bibitem[\protect\citeauthoryear{{Corbel}, {Koerding}  \& {Kaaret}}{{Corbel}
  et~al.}{2008}]{Corbel2008}
{Corbel} S.,  {Koerding} E.,   {Kaaret} P.,  2008, \mn@doi [\mnras]
  {10.1111/j.1365-2966.2008.13542.x}, \href
  {http://adsabs.harvard.edu/abs/2008MNRAS.389.1697C} {389, 1697}

\bibitem[\protect\citeauthoryear{{Esin}, {McClintock}  \& {Narayan}}{{Esin}
  et~al.}{1997}]{Esin1997}
{Esin} A.~A.,  {McClintock} J.~E.,   {Narayan} R.,  1997, \mn@doi [\apj]
  {10.1086/304829}, \href
  {https://ui.adsabs.harvard.edu/abs/1997ApJ...489..865E} {489, 865}

\bibitem[\protect\citeauthoryear{{Falcke}, {K{\"o}rding}  \&
  {Markoff}}{{Falcke} et~al.}{2004}]{F04}
{Falcke} H.,  {K{\"o}rding} E.,   {Markoff} S.,  2004, \mn@doi [\aap]
  {10.1051/0004-6361:20031683}, \href
  {http://adsabs.harvard.edu/abs/2004A%26A...414..895F} {414, 895}

\bibitem[\protect\citeauthoryear{Farr, Sravan, Cantrell, Kreidberg, Bailyn,
  Mandel  \& Kalogera}{Farr et~al.}{2011}]{Farr_2011}
Farr W.~M.,  Sravan N.,  Cantrell A.,  Kreidberg L.,  Bailyn C.~D.,  Mandel I.,
    Kalogera V.,  2011, \mn@doi [The Astrophysical Journal]
  {10.1088/0004-637x/741/2/103}, 741, 103

\bibitem[\protect\citeauthoryear{{Fender}}{{Fender}}{2001}]{Fender:2001}
{Fender} R.~P.,  2001, \mn@doi [MNRAS] {10.1046/j.1365-8711.2001.04080.x},
  \href {http://adsabs.harvard.edu/abs/2001MNRAS.322...31F} {322, 31}

\bibitem[\protect\citeauthoryear{{Fender}, {Maccarone}  \& {Heywood}}{{Fender}
  et~al.}{2013}]{Fender:2013}
{Fender} R.~P.,  {Maccarone} T.~J.,   {Heywood} I.,  2013, \mn@doi [MNRAS]
  {10.1093/mnras/sts688}, \href
  {http://adsabs.harvard.edu/abs/2013MNRAS.430.1538F} {430, 1538}

\bibitem[\protect\citeauthoryear{{Ferri{\`e}re}, {Gillard}  \&
  {Jean}}{{Ferri{\`e}re} et~al.}{2007}]{Ferriere2007}
{Ferri{\`e}re} K.,  {Gillard} W.,   {Jean} P.,  2007, \mn@doi [AAP]
  {10.1051/0004-6361:20066992}, \href
  {http://adsabs.harvard.edu/abs/2007A%26A...467..611F} {467, 611}

\bibitem[\protect\citeauthoryear{Gaggero, Bertone, Calore, Connors, Lovell,
  Markoff  \& Storm}{Gaggero et~al.}{2017}]{Gaggero:2016dpq}
Gaggero D.,  Bertone G.,  Calore F.,  Connors R. M.~T.,  Lovell M.,  Markoff
  S.,   Storm E.,  2017, \mn@doi [Phys. Rev. Lett.]
  {10.1103/PhysRevLett.118.241101}, 118, 241101

\bibitem[\protect\citeauthoryear{{Gallo}, {Fender}  \& {Pooley}}{{Gallo}
  et~al.}{2003}]{Gallo2003}
{Gallo} E.,  {Fender} R.~P.,   {Pooley} G.~G.,  2003, \mn@doi [\mnras]
  {10.1046/j.1365-8711.2003.06791.x}, \href
  {http://adsabs.harvard.edu/abs/gfp03} {344, 60}

\bibitem[\protect\citeauthoryear{{Gallo} et~al.,}{{Gallo}
  et~al.}{2014}]{Gallo2014}
{Gallo} E.,  et~al., 2014, \mn@doi [\mnras] {10.1093/mnras/stu1599}, \href
  {http://adsabs.harvard.edu/abs/2014MNRAS.445..290G} {445, 290}

\bibitem[\protect\citeauthoryear{{He}, {Ricotti}  \& {Geen}}{{He}
  et~al.}{2019}]{HeRG:2019}
{He} C.-C.,  {Ricotti} M.,   {Geen} S.,  2019, \mn@doi [\mnras]
  {10.1093/mnras/stz2239}, \href
  {https://ui.adsabs.harvard.edu/abs/2019MNRAS.489.1880H} {489, 1880}

\bibitem[\protect\citeauthoryear{{He}, {Ricotti}  \& {Geen}}{{He}
  et~al.}{2020}]{HeRG:2020}
{He} C.-C.,  {Ricotti} M.,   {Geen} S.,  2020, \mn@doi [\mnras]
  {10.1093/mnras/staa165}, \href
  {https://ui.adsabs.harvard.edu/abs/2020MNRAS.492.4858H} {492, 4858}

\bibitem[\protect\citeauthoryear{Hobbs, Lorimer, Lyne  \& Kramer}{Hobbs
  et~al.}{2005}]{Hobbs_2005}
Hobbs G.,  Lorimer D.~R.,  Lyne A.~G.,   Kramer M.,  2005, \mn@doi [Monthly
  Notices of the Royal Astronomical Society]
  {10.1111/j.1365-2966.2005.09087.x}, 360, 974–992

\bibitem[\protect\citeauthoryear{{Hong} et~al.,}{{Hong}
  et~al.}{2016}]{Hong:2016}
{Hong} J.,  et~al., 2016, \mn@doi [\apj] {10.3847/0004-637X/825/2/132}, \href
  {http://adsabs.harvard.edu/abs/2016ApJ...825..132H} {825, 132}

\bibitem[\protect\citeauthoryear{{Hoyle} \& {Lyttleton}}{{Hoyle} \&
  {Lyttleton}}{1939}]{Hoyle1939}
{Hoyle} F.,  {Lyttleton} R.~A.,  1939, \mn@doi [Proceedings of the Cambridge
  Philosophical Society] {10.1017/S0305004100021150}, \href
  {http://adsabs.harvard.edu/abs/1939PCPS...35..405H} {35, 405}

\bibitem[\protect\citeauthoryear{Irrgang, Wilcox, Tucker  \&
  Schiefelbein}{Irrgang et~al.}{2013}]{Irrgang_2013}
Irrgang A.,  Wilcox B.,  Tucker E.,   Schiefelbein L.,  2013, \mn@doi
  [Astronomy & Astrophysics] {10.1051/0004-6361/201220540}, 549, A137

\bibitem[\protect\citeauthoryear{{Kruijssen} et~al.,}{{Kruijssen}
  et~al.}{2019}]{2019MNRAS.484.5734K}
{Kruijssen} J.~M.~D.,  et~al., 2019, \mn@doi [\mnras] {10.1093/mnras/stz381},
  \href {https://ui.adsabs.harvard.edu/abs/2019MNRAS.484.5734K} {484, 5734}

\bibitem[\protect\citeauthoryear{{Lazio} \& {Cordes}}{{Lazio} \&
  {Cordes}}{2008}]{2008ApJS..174..481L}
{Lazio} T. J.~W.,  {Cordes} J.~M.,  2008, \mn@doi [\apjs] {10.1086/521676},
  \href {https://ui.adsabs.harvard.edu/abs/2008ApJS..174..481L} {174, 481}

\bibitem[\protect\citeauthoryear{{Maccarone}}{{Maccarone}}{2005}]{Maccarone:2005}
{Maccarone} T.~J.,  2005, \mn@doi [Monthly Notices of the Royal Astronomical
  Society] {10.1111/j.1745-3933.2005.00039.x}, \href
  {http://adsabs.harvard.edu/abs/2005MNRAS.360L..30M} {360, L30}

\bibitem[\protect\citeauthoryear{Manshanden, Gaggero, Bertone, Connors  \&
  Ricotti}{Manshanden et~al.}{2019}]{Manshanden:2018tze}
Manshanden J.,  Gaggero D.,  Bertone G.,  Connors R. M.~T.,   Ricotti M.,
  2019, \mn@doi [JCAP] {10.1088/1475-7516/2019/06/026}, 1906, 026

\bibitem[\protect\citeauthoryear{{Markoff}, {Nowak}  \& {Wilms}}{{Markoff}
  et~al.}{2005}]{Markoff2005}
{Markoff} S.,  {Nowak} M.~A.,   {Wilms} J.,  2005, \mn@doi [\apj]
  {10.1086/497628}, \href
  {https://ui.adsabs.harvard.edu/abs/2005ApJ...635.1203M} {635, 1203}

\bibitem[\protect\citeauthoryear{{Merloni}, {Heinz}  \& {di Matteo}}{{Merloni}
  et~al.}{2003}]{Merloni:2003}
{Merloni} A.,  {Heinz} S.,   {di Matteo} T.,  2003, \mn@doi [\mnras]
  {10.1046/j.1365-2966.2003.07017.x}, \href
  {http://adsabs.harvard.edu/abs/2003MNRAS.345.1057M} {345, 1057}

\bibitem[\protect\citeauthoryear{{Miller-Jones}, {Jonker}, {Maccarone},
  {Nelemans}  \& {Calvelo}}{{Miller-Jones} et~al.}{2011}]{MillerJones2011}
{Miller-Jones} J.~C.~A.,  {Jonker} P.~G.,  {Maccarone} T.~J.,  {Nelemans} G.,
  {Calvelo} D.~E.,  2011, \mn@doi [\apjl] {10.1088/2041-8205/739/1/L18}, \href
  {http://adsabs.harvard.edu/abs/2011ApJ...739L..18M} {739, L18}

\bibitem[\protect\citeauthoryear{{Narayan} \& {Yi}}{{Narayan} \&
  {Yi}}{1994}]{Narayan:1994}
{Narayan} R.,  {Yi} I.,  1994, \mn@doi [ApJ Letters] {10.1086/187381}, \href
  {http://adsabs.harvard.edu/abs/1994ApJ...428L..13N} {428, L13}

\bibitem[\protect\citeauthoryear{Park \& Ricotti}{Park \&
  Ricotti}{2011}]{Park:2010yh}
Park K.,  Ricotti M.,  2011, \mn@doi [Astrophys. J.]
  {10.1088/0004-637X/739/1/2}, 739, 2

\bibitem[\protect\citeauthoryear{Park \& Ricotti}{Park \&
  Ricotti}{2012}]{Park:2011rf}
Park K.,  Ricotti M.,  2012, \mn@doi [Astrophys. J.]
  {10.1088/0004-637X/747/1/9}, 747, 9

\bibitem[\protect\citeauthoryear{Park \& Ricotti}{Park \&
  Ricotti}{2013}]{Park:2012cr}
Park K.,  Ricotti M.,  2013, \mn@doi [Astrophys. J.]
  {10.1088/0004-637X/767/2/163}, 767, 163

\bibitem[\protect\citeauthoryear{{Pellegrini}}{{Pellegrini}}{2005}]{Pellegrini:2005}
{Pellegrini} S.,  2005, \mn@doi [ApJ] {10.1086/429267}, \href
  {http://adsabs.harvard.edu/abs/2005ApJ...624..155P} {624, 155}

\bibitem[\protect\citeauthoryear{{Perna}, {McDowell}, {Menou}, {Raymond}  \&
  {Medvedev}}{{Perna} et~al.}{2003}]{Perna:2003}
{Perna} R.,  {McDowell} J.,  {Menou} K.,  {Raymond} J.,   {Medvedev} M.~V.,
  2003, \mn@doi [ApJ] {10.1086/378855}, \href
  {http://adsabs.harvard.edu/abs/2003ApJ...598..545P} {598, 545}

\bibitem[\protect\citeauthoryear{{Plotkin}, {Markoff}, {Kelly}, {K{\"o}rding}
  \& {Anderson}}{{Plotkin} et~al.}{2012a}]{2012MNRAS.419..267P}
{Plotkin} R.~M.,  {Markoff} S.,  {Kelly} B.~C.,  {K{\"o}rding} E.,   {Anderson}
  S.~F.,  2012a, \mn@doi [\mnras] {10.1111/j.1365-2966.2011.19689.x}, \href
  {http://adsabs.harvard.edu/abs/2012MNRAS.419..267P} {419, 267}

\bibitem[\protect\citeauthoryear{Plotkin, Markoff, Kelly, K\"{o}rding  \&
  Anderson}{Plotkin et~al.}{2012b}]{Plotkin:2012}
Plotkin R.~M.,  Markoff S.,  Kelly B.~C.,  K\"{o}rding E.,   Anderson S.~F.,
  2012b, \mn@doi [MNRAS] {10.1111/j.1365-2966.2011.19689.x}, 419, 267

\bibitem[\protect\citeauthoryear{{Plotkin}, {Gallo}, {Markoff}, {Homan},
  {Jonker}, {Miller-Jones}, {Russell}  \& {Drappeau}}{{Plotkin}
  et~al.}{2015}]{Plotkin2015}
{Plotkin} R.~M.,  {Gallo} E.,  {Markoff} S.,  {Homan} J.,  {Jonker} P.~G.,
  {Miller-Jones} J. C.~A.,  {Russell} D.~M.,   {Drappeau} S.,  2015, \mn@doi
  [\mnras] {10.1093/mnras/stu2385}, \href
  {https://ui.adsabs.harvard.edu/abs/2015MNRAS.446.4098P} {446, 4098}

\bibitem[\protect\citeauthoryear{{Russell} et~al.,}{{Russell}
  et~al.}{2013}]{Russell2013}
{Russell} D.~M.,  et~al., 2013, \mn@doi [\mnras] {10.1093/mnras/sts377}, \href
  {https://ui.adsabs.harvard.edu/abs/2013MNRAS.429..815R} {429, 815}

\bibitem[\protect\citeauthoryear{{Samland}}{{Samland}}{1998}]{1998ApJ...496..155S}
{Samland} M.,  1998, \mn@doi [\apj] {10.1086/305368}, \href
  {https://ui.adsabs.harvard.edu/abs/1998ApJ...496..155S} {496, 155}

\bibitem[\protect\citeauthoryear{Sanders, Smith, Evans  \& Lucas}{Sanders
  et~al.}{2019}]{Sanders_2019}
Sanders J.~L.,  Smith L.,  Evans N.~W.,   Lucas P.,  2019, \mn@doi [Monthly
  Notices of the Royal Astronomical Society] {10.1093/mnras/stz1630}, 487,
  5188–5208

\bibitem[\protect\citeauthoryear{Shapiro \& Teukolsky}{Shapiro \&
  Teukolsky}{1983}]{Shapiro:1983du}
Shapiro S.~L.,  Teukolsky S.~A.,  1983, {Black holes, white dwarfs, and neutron
  stars: The physics of compact objects}

\bibitem[\protect\citeauthoryear{Sofue}{Sofue}{2013}]{Sofue_2013}
Sofue Y.,  2013, \mn@doi [Publications of the Astronomical Society of Japan]
  {10.1093/pasj/65.6.118}, 65, 118

\bibitem[\protect\citeauthoryear{{Sugimura} \& {Ricotti}}{{Sugimura} \&
  {Ricotti}}{2020}]{SugimuraR:2020}
{Sugimura} K.,  {Ricotti} M.,  2020, \mn@doi [\mnras] {10.1093/mnras/staa1394},
  \href {https://ui.adsabs.harvard.edu/abs/2020MNRAS.495.2966S} {495, 2966}

\bibitem[\protect\citeauthoryear{Tsuna \& Kawanaka}{Tsuna \&
  Kawanaka}{2019}]{Tsuna:2019kny}
Tsuna D.,  Kawanaka N.,  2019, \mn@doi [Mon. Not. Roy. Astron. Soc.]
  {10.1093/mnras/stz1809}, 488, 2099

\bibitem[\protect\citeauthoryear{Tsuna, Kawanaka  \& Totani}{Tsuna
  et~al.}{2018}]{Tsuna:2018oqt}
Tsuna D.,  Kawanaka N.,   Totani T.,  2018, \mn@doi [Mon. Not. Roy. Astron.
  Soc.] {10.1093/mnras/sty699}, 477, 791

\bibitem[\protect\citeauthoryear{Weltman et~al.}{Weltman
  et~al.}{2020}]{Bull:2018lat}
Weltman A.,  et~al., 2020, \mn@doi [Publ. Astron. Soc. Austral.]
  {10.1017/pasa.2019.42}, 37, e002

\bibitem[\protect\citeauthoryear{{Yuan}, {Quataert}  \& {Narayan}}{{Yuan}
  et~al.}{2003}]{yqn03}
{Yuan} F.,  {Quataert} E.,   {Narayan} R.,  2003, \mn@doi [\apj]
  {10.1086/378716}, \href {http://adsabs.harvard.edu/abs/2003ApJ...598..301Y}
  {598, 301}

\bibitem[\protect\citeauthoryear{Özel, Psaltis, Narayan  \& McClintock}{Özel
  et~al.}{2010}]{_zel_2010}
Özel F.,  Psaltis D.,  Narayan R.,   McClintock J.~E.,  2010, \mn@doi [The
  Astrophysical Journal] {10.1088/0004-637x/725/2/1918}, 725, 1918–1927

\makeatother
\end{thebibliography}

\end{document}